\documentclass[11pt]{article}
\usepackage[english]{babel}

\usepackage{amsmath}
\usepackage{amssymb}

\usepackage{mathrsfs}

\usepackage{graphicx}
\usepackage{multicol}
\usepackage{xspace}

\usepackage[a4paper,total={6.2in, 9.3in}]{geometry}

\usepackage[bookmarks=true,bookmarksopen=true,colorlinks=true,breaklinks=true,linkcolor=blue,citecolor=red]{hyperref}

\usepackage[font=small,format=plain,labelfont=bf,up,textfont=it,up]{caption} 

\usepackage[affil-it,max2]{authblk}
\usepackage{enumerate}   

\usepackage{cite}

\usepackage{url}

\newcommand{\Eqref}[1]{Eq.~\eqref{#1}}
\newcommand{\Eqsref}[1]{Eqs.~\eqref{#1}}
\newcommand{\Figref}[1]{Fig.~\ref{#1}}

\newcommand{\Sectionref}[1]{section~\ref{#1}}

\usepackage{amsthm} 
\newtheorem{definition}{Definition}

\newcommand{\s}{\hspace{0.1cm}} 
\newcommand{\St}{\ensuremath{\mathbb{S}^2}\xspace} 
\newcommand{\df} {\mathrm{d}}

\graphicspath{{plots/}}

\title{Spectral-infinite element method approach for computing asymptotically flat initial data sets in general relativity} 
\author[1]{Leon Escobar-Diaz}
\author[2]{Paula Bran}
\affil[1,2]{Department of Mathematics, Universidad del Valle}

\begin{document}

\linespread{1.0} \normalsize


\maketitle

\begin{abstract}
In this work, we introduce a spectral-infinite element method for solving Einstein's constraint equations in hyperbolic form. As an application of this method, we compute asymptotically flat perturbations of a Kerr black hole with small angular momentum. Our numerical framework relies on utilizing a spin-weighted spherical harmonic transform in combination with an infinite element method for solving partial differential equations in unbounded domains.\\



\end{abstract}

\section{Introduction}\label{sec:introduction}

In the theory of general relativity, the formulation of the initial value problem results in two sets of equations that must be simultaneously solved: the evolution and constraint equations. Due to the nonlinear nature of both equation sets, it is widely recognized that their numerical and analytical treatment is generally a complex task (for an in-depth discussion, refer to \cite{lehner2001numerical}).

Presently, there exist several methods in the literature for numerically integrating the evolution equations. Among these methods, the BSSN and generalized harmonic formulations are most commonly employed by the community. Conversely, there are only a few numerical approaches available for solving the constraint equations, which, among other factors, determine the initial data for the problem. For a more detailed discussion on this matter, refer to, for instance, \cite{cook2000initial,shibata2015numerical}.

One of the standard approaches for numerically addressing constraints relies on the  conformal method. This method involves reducing the Hamiltonian and momentum constraints into a set of coupled nonlinear partial differential equations of the elliptic type (for a detailed presentation, refer to \cite{baumgarte2010numerical}). Theoretically, this method allows the determination of any conceivable initial data. However, practical numerical solutions for such systems are intricate due to the nonlinear nature of the equations and the inclusion of boundary conditions.

To alleviate the complexity of the constraint equations, certain conditions can be imposed on the desired solutions. An illustrative example is the \textit{constant mean curvature condition}, which facilitates the decoupling of the Hamiltonian and momentum constraint equations. Indeed, numerous theorems regarding the existence, non-existence, and uniqueness of initial data rely on this simplification (see, for instance, \cite{bartnik2004constraint,isenberg2014initial} for a comprehensive review).

Another simplification strategy is the well-known Bowen-York proposal. Widely used in determining initial data for most current binary black hole simulations, this approach simplifies the elliptic system by assuming that the initial data is conformally flat (refer to \cite{alcubierre2008introduction} for details).

Despite the analytical convenience of employing certain simplification strategies, they inherently limit the range of initial data that can be constructed, thus excluding many intriguing scenarios. For instance, due to Garat and Price's proof of the non-existence of conformally flat slices in the Kerr spacetime (as outlined in \cite{garat2000nonexistence}), obtaining conformally flat initial data sets describing a rotating black hole is unattainable. Notably, attempts to use this method for computing initial data for the Kerr spacetime result in unwanted spurious gravitational wave content lacking clear physical interpretation (refer to, for instance, \cite{alcubierre2008introduction,cook2000initial}).

Such limitations within existing methods have spurred researchers in the last decade to explore alternative approaches for solving the constraint equations. Notably, some significant approaches proposed are based on the concept introduced by Matzner, Huq, and Shoemaker in \cite{matzner1998initial}, involving the use of superposed boosted Kerr black holes in a Kerr–Schild form. For a comprehensive discussion, readers can refer to \cite{cook2000initial}.

Recently, Rácz introduced a novel approach for solving the constraint equations in a series of papers (\cite{racz2014cauchy,racz2014bianchi,racz2015constraints}), which avoids reducing the equations to an elliptic system of differential equations. Depending on the freely assigned components of the initial data, the Hamiltonian and momentum constraints are transformed into either a parabolic-hyperbolic system or a hyperbolic-algebraic system, requiring data solely on a spatial two-dimensional surface. Subsequently, Beyer et al. explored this formulation numerically in \cite{beyer2017asymptotics}, assuming a smoothly foliated three-dimensional initial data surface by a $\rho$-parameter family of topological two-spheres $S_{\rho}$. Utilizing a pseudo-spectral method based on spin-weighted spherical harmonics and an explicit adaptive step-size Runge-Kutta method, they successfully solved the constraint equations as an initial value problem, reproducing a Kerr black hole initial data set.

However, as highlighted in \cite{beyer2017asymptotics}, a major drawback of this implementation is the inability to control the solution's asymptotic behavior or its boundary values near infinity. This limitation arises due to the numerical infrastructure employing a recursive explicit Runge-Kutta scheme for integrating the solutions with the parameter $\rho$, which labels the two-spheres $S_{\rho}$. Consequently, it only provides an approximate solution up to a finite value of $\rho$,  thereby compromising the ability to control both the solutions asymptotic behavior and their boundary data set at infinity. Furthermore, typical of evolution-type problems, numerical instabilities due to evolution integrators affect the accuracy of the results for large values of $\rho$.

In this study, we aim to enhance the pseudo-spectral method introduced in \cite{beyer2017asymptotics} by substituting the Runge-Kutta scheme with an infinite-element scheme. This modification will facilitate solving Rácz's formulation of the constraint equations across the entire domain, enabling control over the solutions' asymptotic decay and their boundary data at infinity. As an application, we will utilize this new numerical approach to discover asymptotically flat perturbations of a rotating black hole with small angular momentum.

The structure of this work is as follows: In Section 2, we provide an overview of Rácz's hyperbolic-algebraic formulation of the Einstein constraint equations and briefly summarize the pseudo-spectral method employed in \cite{beyer2017asymptotics} for numerical solutions. In Section 3, we introduce a spectral-infinite element method for solving partial differential equations in unbounded domains of the form $[a,\infty) \times \mathbb{S}^2$ with $a>0$. Following this, in Section 4, we employ this method in the hyperbolic formulation of the constraint equations to compute asymptotically flat linear perturbations of a Kerr black hole with small angular momentum. Finally, in Section 5, we conclude.

\section{The hyperbolic form of the Einstein constraint equations}\label{section:segunda_seccion}

In this section, we revisit the hyperbolic-algebraic formulation of the constraint equations introduced by Rácz in \cite{racz2015constraints,racz2014bianchi}. The fundamental concept involves conducting a $2+1$ decomposition over the manifold where the Einstein constraint equations are defined.

Throughout this work, abstract tensor indices will be represented by Latin characters $a,b,c,...$, while Greek characters $\mu,\nu,...$ will signify tensor components concerning a specific frame in a $3$-dimensional manifold, considering indices from $0$ to $2$. Moreover, coordinate frame vectors will be denoted as $\partial^a_{\mu}$. Therefore, vectors will be expressed concerning this basis as $v^a = v^{\mu} \partial^a_{\mu}$, utilizing Einstein's summation convention (as described in \cite{Wald:1984un}). 

\subsection{The constraint equations}
Initially, we consider a smooth $3$-dimensional manifold $\Sigma$ equipped with a Riemannian metric $h_{ab}$ and a second fundamental form $\chi_{ab}$ concerning a Lorentzian smooth $4$-dimensional manifold $M$. Essentially, $\Sigma$ is embedded within $M$. 

\begin{definition}
	The pair $ (h_ {ab}, \chi_ {ab})$ represents an initial data for the evolution Einstein equations in vacuum defined on the manifold $M$ if the following tensorial equations on $\Sigma$ are satisfied:  
	\begin{eqnarray}
	R + \chi - \chi_{ab} \chi^{ab} = 0,\label{StandartConstraints1}\\
	\nabla^a \chi_{a b} - \nabla_b \chi =0,\label{StandartConstraints2}
	\end{eqnarray}
	where $h^{ab}$ is the inverse of $h_{ab}$, $\chi := h^{ab} \chi_{ ab} $ is the mean curvature of $\Sigma$ with respect  to $M$, $R $ is the intrinsic curvature of $\Sigma$, and $ \nabla_ a$  is the covariant derivative operator compatible with $h_{ab}$.
\end{definition}
These equations are commonly recognized in the literature as the Einstein constraint equations.

A crucial geometric condition concerning the initial data $(h_{ab}, \chi_{ab})$ is that of asymptotic flatness, which asserts that at substantial distances from a specific region, the spacetime's geometry closely resembles that of Minkowski spacetime. The most comprehensive and modern coordinate-independent definition of asymptotic flatness was introduced by Geroch in \cite{geroch1972structure}. However, for the purposes of this work, we will employ the coordinate-dependent definition outlined in \cite{dain2001asymptotically}, as it aligns more suitably with our objectives.

\begin{definition}\label{def:definition_asymptotically_flat}
An initial data set $(h_{ab},\chi_{ab})$ is an \textit{asymptotically flat initial data}  if there exist coordinates $\{\tilde{x}^{\bar\sigma}\}$ such that their components $(h_{\bar\mu \bar\nu},\chi_{\bar\mu \bar\nu} )$  in the coordinate frame satisfy
 \begin{eqnarray}
  h_{\bar\mu \bar\nu}    &=& (1+\dfrac{2 M}{r}) \delta_{\mu\nu} + \mathcal{O}(r^{-2}), \label{eq:asymptotically_flat_condition1}\\
  \chi_{\bar\mu \bar\nu} &=& \mathcal{O}(r^{-2}),\label{eq:asymptotically_flat_condition2}
\end{eqnarray}
where
\begin{equation*}\label{AsymtoticallyFlatmetric}
r := \sum \limits_{i=1}^{3} (\tilde{x}^{\bar \sigma})^2,
\end{equation*}
$\delta_{\bar\mu \bar\nu}$ denotes the Euclidean metric components with respect to the given coordinate system, and $M$ is some given constant.
\end{definition}
Note that in this definition we have used the $\mathcal{O}$ notation that states the equivalence
\begin{equation}\label{ec:orden_decaimiento}
	f = \mathcal{O}(r^n)  \iff  |f| \leq C r^{n}. 
\end{equation}
for some positive real constant $C$.

\subsection{2+1 splitting of the constraint equations}\label{sec:TheSytemOfEquations}

In what follows, we conduct the $2+1$ splitting of the constraint, which we divide into four steps.   

\subsubsection{Step 1: Folitation of the manifold}
We assume  that the topology of $\Sigma$ admits a complete foliation of topological spheres  $S_{\rho}$, parameterized by level surfaces of a smooth, positive and monotone increasing function $ \rho: \Sigma \to \mathbb{R}^+ $, i.e., we choose the foliation such that 

\begin{equation*}  
\Sigma = \bigcup\limits_{\rho=0}^{\infty}  S_\rho , \quad S_{\rho_i} \cap S_{\rho_j} = \emptyset .
\end{equation*}

\begin{figure}[t] 
	\centering
 	\includegraphics[scale=0.43]{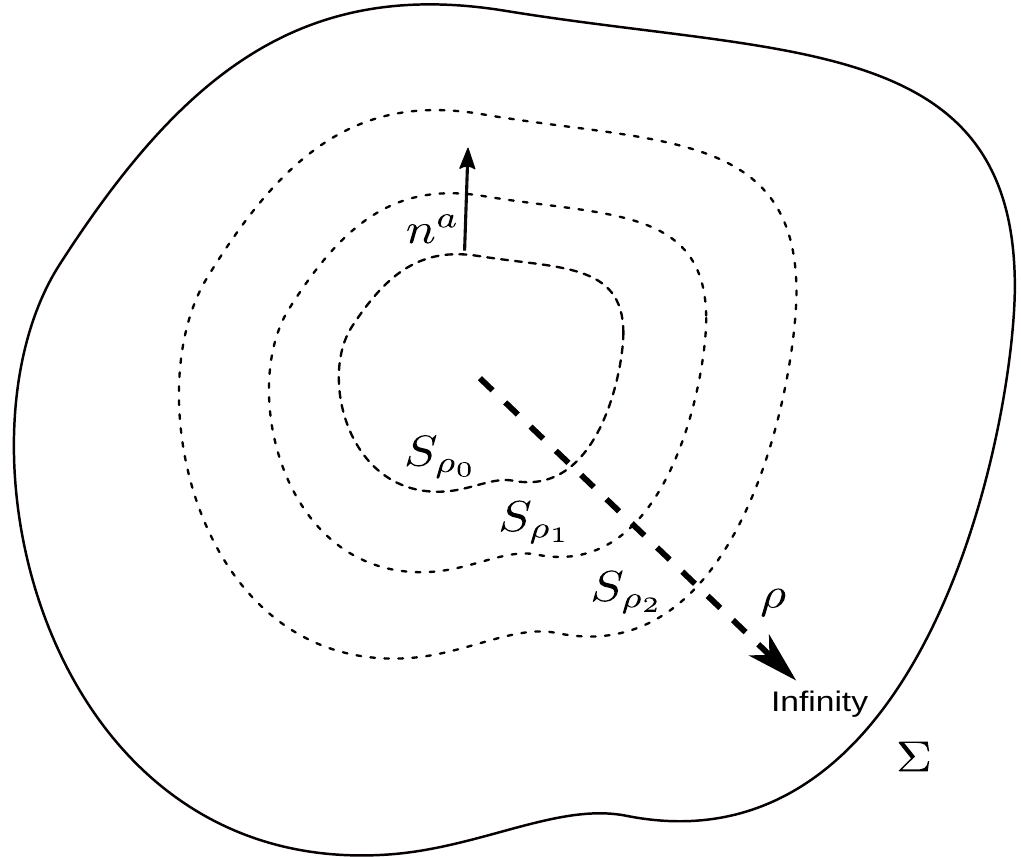}  
	\caption{Foliation scheme of $\Sigma$ with surfaces $S_{\rho_i}$ for  $\rho_i=\rho_0, \rho_1, \rho_2,....$.} \label{fig:figure0} 
\end{figure} 

By topological spheres, we imply that the $S_{\rho_i}$ are conformal to the standard two-sphere $\mathbb{S}^2$, denoted as $S_\rho \simeq \mathbb{S}^2$. To provide clarity, we depict this foliation in \Figref{fig:figure0}. It is important to note that the vector $n^a$ denotes the unit normal vector to the topological spheres $S_\rho$. Following the analogy with the standard $3+1$ decomposition of a spacetime (refer, for instance, to \cite{baumgarte2010numerical}), we select $\rho^{a}$ as the tangent vector to the curves generated by the parameter $\rho$, ensuring it satisfies the relation:
\begin{equation*}
 \rho^a \nabla_a \rho := 1.
\end{equation*}

From the above, it follows that the unitary normal vector to each surface $S_\rho $ can be expressed as 
\begin{equation*}\label{eqc:1}
n^{a} =   \alpha^{-1}  \left (\rho^{a} - \beta^{a} \right),
\end{equation*}
where   $\alpha$ y $\beta^a $ 
will be called the ``\textit{lapse function}'' and  the ``\textit{shift vector}'' associated to the vector $ \rho^a $ with respect to the surfaces $S_{\rho}$.

\subsubsection{Step 2: Decomposition of metric and extrinsic curvature}
Let us consider the  operator
\begin{equation}\label{ec:proyector_equation}
\gamma^{a}_{\ b} := \delta^{a} _ {\ b} - n^{a} n_{b},
\end{equation}
where $\delta^{a} _ {\ b}$ is the standard Kronecker delta. It can be easily proved that it  projects tensors from the manifold $\Sigma$ to the surface $S_\rho$. Furthermore, it induces a  metric on $S_\rho$ as
(see\cite{Wald:1984un})
\begin{equation}\label{eqc:2}
\gamma_{ab} = h_{ab} - n_a \otimes n_b,
\end{equation}
and the convariant derivative $D_a$ compatible with $\gamma_{ab}$ given by
\begin{equation*}
D_a = \gamma_{a}^{\ b} \nabla_b \ .
\end{equation*}
Note that \Eqref{eqc:2} defines a decomposition of the metric $h_{ab}$ in terms of $n_{a}$ and $\gamma_{ab}$.  We can decompose the second fundamental form  $\chi_{ab}$  in terms of   $\alpha $, $\beta^a $ and $\gamma_{ab} $ as follows
\begin{equation}\label{ec:mean_curvature_decomposition}
\chi_{ab} = Z n_{a}\otimes n_{b} + n_{a} \otimes Y_{b} +  n_{b} \otimes Y_{a} + K_{ab} ,
\end{equation}
with
\begin{equation*}\label{ecs:formulae_of_decomposition}
Z := n^{a}  n^{b} \chi_{ab}, \quad Y_{a} := \gamma^{b}_{\ a } n^{c} \chi_{bc}, \quad K_{ab}:= \gamma^{c}_{\ a} 
\gamma^{d}_{\ b} \chi_{cd}.
\end{equation*}
Additionally, the tensor $K_{ab}$ can be expressed in terms of its trace $X:=\gamma^{ab} K_{ab} $, and its trace free part $\mathring{K}$ by 
\begin{equation*}\label{eq:trace_decomposition}
	K_{ab} :=  \mathring{ K } _{ab} + \dfrac{1}{2} \gamma_{ab} X.
\end{equation*}

\subsubsection{Step 3: Decomposition of the constraints equations}\label{sec:decompositionconstraints}

By direct substitution of \Eqref{ec:mean_curvature_decomposition} into \Eqref{StandartConstraints1}, we can express the Hamiltonian constraint in terms of the quantities $Z, X, Y_a$ as follows
\begin{equation}\label{eq:Constraint3}
Z = \dfrac{1}{2 X } \left( 2 Y^a Y_a - \dfrac{  X ^2 }{2}  - \kappa_{0} \right),
\end{equation}
where 
\begin{equation}
\kappa_{0} =  R - \mathring{ K } _{ab} \mathring{ K }^{ab}.
\end{equation}
Similarly, replacing \Eqsref{ec:mean_curvature_decomposition} into the momentum constraints \Eqref{StandartConstraints2} leads, after some computations, to the following expressions (see \cite{racz2015constraints} for details of the procedure)
\begin{equation}\label{eq:Constraint4}
\mathscr{L}_{n} X - D^a Y_a  + 2 \dot{n}^b Y_{b} - \left( Z - \dfrac{1}{2} X \right) H +   \mathring{ K } _{ab}  H^{ab} =0,  
\end{equation}
\begin{equation}\label{eq:Constraint5}
\mathscr{L}_{n} Y_{a}  + \dfrac{1}{ X } \left( Z D_a X - 2  Y^{b} D_{a} Y_{b} + \dfrac{1}{2} D_a ( R - \mathring{ K } _{cd} \mathring{ K }^{cd}  )  \right)  
+ H Y_{a} + \left( Z - \dfrac{1}{2} X \right) \dot{ n }_a + ( D^b - \dot{ n }^b ) \mathring{ K } _{ba}  = 0,    
\end{equation}   

where  the operator $\mathscr{L}_{n}$ denotes the Lie derivative along the normal vector $n^a$,  $\dot{n}_a := n^b D_b n_a = - D_a (\text{ln} \alpha)$, and 
$H_{ab}$ is the second fundamental form of the topological spheres $S_{\rho}$ with respect to $\Sigma$. This tensor, and its trace, are given in terms of $\gamma_{ab}$ and $n^{a}$, respectively, as 
\begin{equation}\label{ec:segundaformaenspheres}
H_{ab} =  \dfrac{1}{2} \mathscr{L}_{n} \gamma_{ab} ,  \quad H:= \gamma^{ab} H_{ab} .
\end{equation}
Note that the above form of the constraints equations (\Eqsref{eq:Constraint3}, (\ref{eq:Constraint4}) and (\ref{eq:Constraint5})) requires that $X \ne 0$. This condition  implies that time-symmetric initial data (initial data such that $\chi_{ab}$ vanishes) cannot be obtained out of these equations.

\subsubsection{The constraints as an initial value problem}

To represent the new form of the constraint equations as an initial value problem, we initially opt for coordinates that align with the foliation. Hence, the points on $\Sigma$ will be denoted as $(\rho, x^{\mu})$, where $x^{\mu}$ represent coordinates on $S_\rho$. Subsequently, we will employ $\mu, \nu, ...$ to denote tensor indices on $S_{\rho}$, taking values from $1$ to $2$.

Following this, we select $\rho^{a}$ as the frame vector $\partial^a_{0}$. In these coordinates, it is evident that $\partial^a_{0}:=\partial_{\rho}$. Thus, utilizing the frame $(\partial^a_{0}, \partial^{a}{\mu})$, where $\partial^{a}{\mu}$ signifies derivatives along the coordinates parameterizing the topological spheres $S_{\rho}$, we can express the Lie derivatives of $X$ and the components of $Y_{a}$ as follows:
\begin{equation*}
\mathscr{L}_{n} X =  \alpha^{-1} (\partial_{\rho} X  - \beta^{ \nu} \partial_{ \nu} X ),
\end{equation*}
\begin{equation*}
\mathscr{L}_{n} Y_{ \mu} = \alpha^{-1} ( \partial_{\rho} Y_{ \mu} - \beta^{  \nu } \partial_{  \nu } Y_{ \mu} -  Y_{ \nu} \partial_{ \mu}     \beta^{  \nu } ).
\end{equation*}
Note that because of the tensors $Y_{a},\beta^{a}, \gamma_{ab}, \mathring{K}_{ab}$ are completely in $S_{\rho}$, their components with respect to $\partial^a_{0}$ must vanish, i.e., $Y_{0}=\beta^{0}= \gamma_{00}= \gamma_{0   \mu} = \mathring{K}_{00}=\mathring{K}_{0  \mu} = 0$. Hence, by a direct substitution of the above into \Eqsref{eq:Constraint4} and (\ref{eq:Constraint5}), we can write the constraint equations in the following  matricial form 
\begin{gather}\label{ec:matricial_system}
\partial_{\rho}  \begin{bmatrix}    X \vspace{0.2cm}   \\ Y_{ \mu} \vspace{0.2cm}  \end{bmatrix} 
=
\begin{bmatrix}
\beta^{ \sigma} \partial_{ \sigma}  & & \alpha \gamma^{\nu  \sigma} \partial_{ \sigma} \\
-\dfrac{\alpha Z }{X} \partial_{ \mu}  & & \left( \beta^{\sigma} +\dfrac{2 \alpha \gamma^{\nu   \eta} Y_{\eta} \delta^{\ \sigma}_{\mu} }{X} \right)  \partial_{\sigma} \\
\end{bmatrix}
\begin{bmatrix}
X \vspace{0.2cm} \\ Y_{ \nu} \vspace{0.2cm} 
\end{bmatrix}
+
\begin{bmatrix}
F( X, Y_{ \mu}, \alpha , \beta^{ \mu} ,\mathring{K}_{\mu \nu}  ) \vspace{0.2cm}  \\
G_\mu( X, Y_{ \mu}, \alpha , \beta^{ \mu}  ,\mathring{K}_{ \mu  \nu} )  \vspace{0.2cm} 
\end{bmatrix}   ,
\end{gather}
where the functions $F$ and $G_{ \mu}$ do not contain any derivative of either  $X$ or $Y_{\mu}$. 
Choosing $\hat{u}$ as the vector $\hat{u}:=[ X \, ,  Y_{ \mu } ]$, we can write the above matricial system \Eqref{ec:matricial_system} in the compact form 
\begin{equation}\label{ecuacionevoluion}
\partial_\rho \hat{u} = L \hat{u} + A, 
\end{equation}
where $L$ is the operator (matrix) that contains all the derivatives over the components of $\hat{u}$, and $A$ is a vector composed by $[F , G_\mu ]$.  Note that if we substitute the algebraic condition  \Eqref{eq:Constraint3} into the system \Eqref{ec:matricial_system}, we can remove the variable $Z$ from the system. Hence, it is clear that the solution of the Einstein constraints in this form is completely determined by the vector $\tilde u$ whenever the metric $g_{ab}$ on $\Sigma$ and the decomposition quantities that define the foliations $S_\rho$ are given; namely, $\alpha, \beta^a$ and $\mathring{K}_{ab}$.

In \cite{racz2015constraints, racz2014bianchi}, Rácz established that if the inequality $ZX < 0$ holds for all $\rho$ within a certain interval $[\rho_0,\rho_f]$, then the system described in \Eqref{ecuacionevoluion} constitutes a first-order hyperbolic system of PDEs in the variables $X$ and $Y_{\mu}$. This condition ensures the local existence and uniqueness of the solution within the interval $[\rho_0,\rho_f]$ given some initial data $u$ at $S_{\rho_0}$. Consequently, in this context, the system \Eqref{ecuacionevoluion} defines an initial value problem concerning the variable $\rho$.

\subsection{A pseudo-spectral approach for solving the constraints}

Recently, in \cite{beyer2017asymptotics}, Beyer et al. achieved the first numerical solutions of the system described in \Eqref{ecuacionevoluion} by employing a pseudo-spectral method based on spin-weighted spherical harmonics (\textbf{SWSH}) and an explicit adaptive step-size Runge-Kutta method (for comprehensive details on Runge-Kutta methods, refer to \cite{butcher2008numerical}). Specifically, their work successfully numerically solved the system \Eqref{ecuacionevoluion} as an initial value problem for any prescribed initial data $u$ at some $S_{\rho_0}$. This advancement enabled the replication of known solutions of the constraint equations, such as the Kerr black hole initial data in Kerr-Schild coordinates. The following section provides a brief overview of this implementation.

\subsubsection{The spin-weighted spherical harmonics and the eths operators}\label{sec:swsh_seccion1}

To commence, we introduce standard spherical coordinates $(\theta,\varphi)$ on $\mathbb{S}^2$. Due to the equivalence $S_\rho \simeq \mathbb{S}^2$, functions defined on $S_\rho$ can be considered simply as functions defined on $\mathbb{S}^2$. Let $f: \mathbb{S}^2 \to \mathbb{C}$ be a square-integrable complex function, denoted as $f\in L^2(\mathbb{S}^2)$ (for a formal presentation of $L^p$ spaces, refer to \cite{gilbarg2015elliptic}). As per Penrose and Rindler \cite{penrose1984spinors}, we assign spin-weight $s$ to $f$ if it transforms under the action of the one-parameter group U(1) within the tangent plane at every point $(\theta,\varphi) \in \St$ as $f \to e^{\text{i} s \zeta} f$, where $\zeta$ denotes the group parameter. Moreover, $f$ can be expressed as
\begin{equation}\label{ec:spectral_decomposition}
f = \sum\limits_{l,m}  \hspace{0.1cm}_{s}a_{lm} \hspace{0.2cm}_{s}Y_{lm} (\theta,\varphi) :=  \sum\limits_{l=|s|}^{\infty}  \sum\limits_{m=-l}^{l} \hspace{0.1cm}_{s}a_{lm} \hspace{0.2cm}_{s}Y_{lm} (\theta,\varphi) , 
\end{equation}
where $_{s}Y_{lm}( \theta , \varphi)$ are the \textbf{swsh} which satisfy the following relation 
\begin{eqnarray}\label{integral_properties_spherical_harmonics}
  \langle \  {}_{s_1} Y_{l_1 m_1 }(\theta,\varphi) \ , \ _{s_2}\overline{Y}_{l_2 m_2}(\theta,\varphi) \ \rangle &=& \int \limits_{\mathbb{S}^2} \s  {}_{s_1} Y_{l_1 m_1 }(\theta,\varphi) \: _{s_2}\overline{Y}_{l_2 m_2}(\theta,\varphi) \s \sin \theta d\theta d\varphi \nonumber \\
   &=& \delta_{l_1 l_2} \delta_{m_1 m_2} \delta_{s_1 s_2}.
\end{eqnarray}
Note that $\langle \ , \ \rangle $ is just the standard inner product of $L^2 (\mathbb{S}^2)$ induced by the norm.  Furthermore,  the above relation implies that the \textbf{swsh} are a orthogonal basis for any function $f$ with spin weight $s$ defined on $\mathbb{S}^2$. The explicit form of the \textbf{swsh} in terms of the Legendre polynomials can be found, for example, in \cite{alcubierre2008introduction}.

Other important properties, which we include for later use, are the relation between the complex conjugates
\begin{equation}\label{ec:complex_conjugate_swsh}
{ }_{s} \overline{Y}_{lm} = (-1)^{s+m} { }_{-s} Y_{l(-m)},
\end{equation}

and the relation between the triple integration of \textbf{swsh} and the well-known $3j$-symbols (see for example \cite{alcubierre2008introduction}) \begin{equation}\label{ec:clebsh_jordan_formula}
\begin{aligned}
\int \limits_{\mathbb{S}^2} \s  {}_{s_1} Y_{l_1 m_1 }(\theta,\varphi) \: { }_{s_2} Y_{l_2 m_2}(\theta,\varphi) \: {}_{s_3} Y_{l_3 m_3 }(\theta,\varphi) \s d\Omega = \quad \\
  \sqrt{ \dfrac{ (2 l_1 +1)(2 l_2 +1)(2 l_3 +1) }{4 \pi} }   \begin{pmatrix}
l_1 & l_2 & l_3\\
m_1 & m_2 & m_3
\end{pmatrix} \begin{pmatrix}
l_1 & l_2 & l_3\\
s_1 & s_2 & s_3
\end{pmatrix}.
\end{aligned}
\end{equation}

The \textit{eth-operators}, denoted by $\eth$ and $\bar{\eth}$, are  defined as (see for instance \cite{newman1966note}) 
\begin{equation}\label{eq:def_eths}
\eth f       := \partial_\theta f - \dfrac{\text{i}}{ \text{sin} \theta} \partial_\varphi f- s f \text{cot} \theta, \quad 
\bar{\eth} f := \partial_\theta f + \dfrac{\text{i}}{ \text{sin} \theta} \partial_\varphi f + s f \text{cot} \theta .
\end{equation}
These two operators \textit{rise} and \textit{lower} the spin-weight of the \textbf{swsh} basis by means of the following two properties 
\begin{eqnarray}
\begin{aligned}\label{eq:eths}
\eth  \hspace{0.1cm}_{s}Y_{lm} (\theta,\varphi)  &= S(l,s, \hspace{0.3cm} 1) \hspace{0.1cm}_{s+1}Y_{lm} (\theta,\varphi) ,  \\
\bar{\eth}   \hspace{0.1cm}_{s}Y_{lm} (\theta,\varphi)   &= S(l,s, -1) \hspace{0.1cm}_{s-1}Y_{lm} (\theta,\varphi) , \\
\end{aligned}
\end{eqnarray}
where 
\begin{equation}\label{ec:ecuationforS}
S(l,s, \zeta) := -\zeta \sqrt{ (l -\zeta s)(l+\zeta s+1) }.
\end{equation} 
Note that  if a function $f$ has spin-weight of $s$, then $\eth f$ will have a spin-weight of $s+1$, and $\bar{\eth} f$ will have spin-weight of $s-1$.

\subsubsection{Choosing of the non-coordinate frame}\label{sec:coordinates}

We select the \textit{non-coordinate frame} $(\mathbf{e}^a_1, \mathbf{e}^a_2)$ on $S_{\rho}$ as follows:\footnote{This frame finds extensive usage in the well-known Newman-Penrose formalism for gravitational waves, as demonstrated in, for instance, \cite{alcubierre2008introduction}.}
\begin{equation}\label{eq:referenceframe}
\quad \mathbf{e}^{a}_{1}:=\frac 1{\sqrt 2}\left(\partial^a_{\theta}-\frac{\text{i}} {\sin\theta}\partial^a_\varphi\right), \quad \mathbf{e}^{a}_{2}  :=\frac 1{\sqrt 2}\left(\partial^a_{\theta}+\frac{\text{i}} {\sin\theta}\partial^a_\varphi\right),
\end{equation}
where $\partial^a_{\theta}$ and $\partial^a_{\varphi}$ correspond to the coordinate vectors associated to the standard spherical coordinates. Additionally, we define the coframe $(\mathbf{w}^1_a, \mathbf{w}^2_a)$  such that 
\begin{equation}\label{eq:referencecoframe}
\mathbf{w}^1_a := \frac 1{\sqrt 2}\left(\df \theta_a + \text{i} \sin\theta \ \df \varphi_a \right), \quad 
\mathbf{w}^2_a := \frac 1{\sqrt 2}\left(\df \theta_a - \text{i} \sin\theta \ \df \varphi_a \right),
\end{equation}
where $\df$ is the exterior derivative operator in $S_{\rho}$ (see \cite{nakahara2003geometry}), i.e.,  $\df \theta_a = D_a \theta$ and $\df \varphi_a = D_a \varphi$. Clearly $\mathbf{w}^\nu_a \mathbf{e}^{a}_{\mu} = \delta^{\nu}_{\mu}$. Further, it can be easily proved that the frame vectors transform under a rotation by an angle $\zeta$ in the tangent plane at  every point of $\St$, that is under the action of the U(1) group (see for instance \cite{escobar2016studies}),  as 
\begin{equation}\label{eq:transformationrules_spin}
 \mathbf{e}^a_\mu  \to  e^{\text{i} \Omega_\mu \zeta} \mathbf{e}^a_\mu , \text{ with } \Omega_\mu = \left\{
	\begin{array}{ll}
		\hspace{0.3cm}  1  & \mbox{if } \mu=1 ,\\
		-1 & \mbox{if } \mu=2 .
	\end{array}
\right.
\end{equation}
Since scalar numbers must be invariant under rotation, it follows from \Eqsref{eq:referencecoframe} that  the coframe   must transform under the action of U(1) as $\mathbf{w}^a_\mu  \to  e^{- \text{i} \Omega_\mu \zeta} \mathbf{w}^a_\mu$.  Hence, tensor components with respect that frame (and coframe) must transform  as  
\begin{equation*}
T^{\mu...\nu}_{\qquad  \sigma...\lambda} \to e^{\text{i} s \zeta} T^{\mu...\nu}_{\qquad  \sigma...\lambda},
\end{equation*}
where the spin weight $s$ depends on the number of frame vectors $\mathbf{e}^a_{\mu}$ and coframe covectors $\mathbf{w}^a_\mu$ with respect to the tensor components are taken. Thus, tensor components have a well defined spin weight. For example, 
the components $\gamma_{\mu\nu}$ of the metric tensor $\gamma_{ab}$ transform as
\begin{eqnarray*}
\gamma_{\mu\nu} = \mathbf{e}^a_\mu \mathbf{e}^b_\nu \gamma_{ab} \to 
 e^{\text{i} \Omega_\mu \zeta} e^{\text{i} \Omega_\nu \zeta} \mathbf{e}^a_\mu \mathbf{e}^b_\nu \gamma_{ab} = e^{\text{i} (\Omega_\mu+\Omega_\nu) \zeta} \gamma_{\mu\nu}.
\end{eqnarray*}
This reveals that their spin-weight $s$ is determined by $\Omega_\mu+\Omega_\nu$. It is important to note that this derivation relies on the property that abstract tensors remain unchanged under rotations of the frame.

From the above, and the discussion of \Sectionref{sec:swsh_seccion1}, it clearly follows that  tensor components with spin weight $s$ can be written as  
\begin{equation}\label{eq:TensorfunctionS2}
T^{\mu...\nu}_{\qquad  \sigma...\lambda} =  \sum\limits_{l,m}  \hspace{0.1cm}_{s}a_{lm} \hspace{0.2cm}_{s}Y_{lm} (\theta,\varphi),
\end{equation}
where the constants $_{s}a_{lm}$ are known as the \textit{spectral coefficients}. 

The operation of the frame vectors on tensor components with spin weight $s$ can be computed using the eth operators as follows. Utilizing \Eqref{eq:def_eths} in conjunction with the coordinate definition of the frame vectors in \Eqsref{eq:referenceframe}, it follows that
\begin{eqnarray*}\label{eq:ethm}
\mathbf{e}^a_\tau( T^{\mu...\nu}_{\qquad  \sigma...\lambda}  ) = \dfrac{1}{\sqrt 2} \left(  \eth_\tau T^{\mu...\nu}_{\qquad  \sigma...\lambda}  +  \Omega_\tau  \ s \ \cot\theta  \  T^{\mu...\nu}_{\qquad  \sigma...\lambda}  \right),
\end{eqnarray*}
where we have used the definition
\begin{equation}\label{ec:indiceseths}
\eth_{\tau}:=(\eth,\bar \eth ).
\end{equation}

Moreover, using these relations and the fact that the metric $\gamma_{ab}$ is conformal to that of  the two-sphere $\mathring \gamma_{ab}$, i.e., $\gamma_{ab} = \Theta^2 \mathring \gamma_{ab} $ with $\Theta \neq 0$, it follows  that the components $D_{\tau} T^{\mu...\nu}_{\qquad  \sigma...\lambda}$ of the tensor $ D_a T^{b...c}_{\quad \ \ d...f}$ can be expressed in terms of the eths operators as (see for instance \cite{escobar2016studies}) 
\begin{equation}\label{ec:relation_covariant_derivative_eth}
\begin{split}
  D_{\tau} T^{\mu...\nu}_{\qquad  \sigma...\lambda} &= \dfrac{1}{\sqrt{2}} \eth_\tau T^{\mu...\nu}_{\qquad  \sigma...\lambda}+ \ C^{\mu}_{\tau \omega} T^{\omega...\nu}_{\qquad  \sigma...\lambda} \ + \ ...\ + \ C^{\nu}_{\tau \omega}T^{\mu...\omega}_{\qquad  \sigma...\lambda} \\ 
&  - \ C^{\omega}_{\tau \sigma}T^{\mu...\nu}_{\qquad  \omega...\lambda}- \ ... \ - \ C^{\omega}_{\tau \lambda} T^{\mu...\nu}_{\qquad  \sigma...\omega},
\end{split}
\end{equation} 
where the symbols $C^{i}_{jk}$ are the conformal transformation coefficients given by
\begin{equation}\label{ec:conformalcoefficients}
C^{\lambda}_{\mu \nu} = \dfrac{1}{\sqrt{2}} \Theta^{-1} \mathring{ \gamma }^{\lambda \sigma} \left(  \mathring{ \gamma }_{\nu \sigma} \eth_{\mu} \Theta  + \mathring{ \gamma }_{\mu \sigma} \eth_{\nu} \Theta  - \mathring{ \gamma }_{\mu \nu } \eth_{\sigma} \Theta \right).
\end{equation}
Note that once one knows the spectral decomposition of the tensor components as in    \Eqref{eq:TensorfunctionS2}, one can use the relations of \Eqsref{eq:eths} to find the first term of the right hand side of \Eqref{ec:relation_covariant_derivative_eth} in terms of the \textbf{swsh} as
\begin{equation*}\label{ec:eths_acting_on_components}
 \eth_\tau T^{\mu...\nu}_{\qquad  \sigma...\lambda} = \sum\limits_{l,m}  \hspace{0.1cm}_{s}a_{lm} \ \eth_\tau \hspace{0.2cm}_{s}Y_{lm} (\theta,\varphi) = \sum\limits_{l,m}   \hspace{0.1cm}_{s}a_{lm} \  S(l,s, \Omega_\tau )  \hspace{0.2cm}_{s + \Omega_\tau }Y_{lm} (\theta,\varphi).\\
\end{equation*}
In summary, projecting tensors onto the frame $\mathbf{e}^a_\mu$ allows expressing their components in terms of the \textbf{SWSH}. Consequently, this approach enables the computation of covariant derivatives using the eth-operators. As detailed in the subsequent subsection, this constitutes the fundamental concept behind the pseudo-spectral approach introduced by Beyer et al. \cite{beyer2017asymptotics}, and it will play a pivotal role in our implementation.


\subsubsection{The pseudo-spectral approach}\label{sec:pseudo_spectral_method}
To initiate, we establish standard spherical coordinates $(r,\theta,\varphi)$ in $\Sigma$. The frame $(\mathbf{e}^a_0 ,\mathbf{e}^a_1, \mathbf{e}^a_2)$ is chosen such that $\mathbf{e}^a_0 = \partial^a_r$, while $\mathbf{e}^a_1$ and $\mathbf{e}^a_2$ constitute the frame vectors defined for the surfaces $S_{\rho}$ in \Eqref{eq:referenceframe}. Typically, the smooth function $\rho$ defining the foliations $S_{\rho}$ takes the form $\rho = \rho(r,\theta,\varphi)$. However, for the sake of simplification, we assume that suitable foliations $S_{\rho}$ can be found where $\rho = \rho(r)$, allowing the vector $\partial^{a}{\rho}$ to be expressed as $\partial^{a}{\rho} = \rho^0 \mathbf{e}^a_0$.

By projecting the tensorial equations \Eqref{eq:Constraint4} and \Eqref{eq:Constraint5} onto the frame $(\mathbf{e}^a_0 ,\mathbf{e}^a_1, \mathbf{e}^a_2)$ and utilizing \Eqref{ec:relation_covariant_derivative_eth} to represent the covariant derivatives in terms of the eth operators, we observe that the system \Eqref{ecuacionevoluion}, coupled with the initial data at $S_{\rho_0}$, can be condensed into the following boundary value problem:
\begin{equation}\label{ec:prototipe0}
\begin{split}
\partial_{r} \hat{u} &= F( \hat{u}, \eth \hat{u}, \bar \eth \hat{u}, \hat{B} ),\\
\hat{u}|_{\rho_0} &= \hat{u}_0,
\end{split}
\end{equation}
where  $\hat{u}_0$ is the value of $\hat{u}$ at the initial topological spheres $S_{\rho_0}$, and $B$ contains the remaining quantities that define the $2+1$ foliation on $\Sigma$, i.e., $B= B(\alpha, \beta^\mu, \gamma_{\mu \nu}, \mathring{K}_{\mu \nu})$. Because of $Y_{a}$ lies on the hypersurfaces $S_{\rho}$, it is clear that it can be written as 
\begin{equation*}
Y_{a} = Y_{1} \mathbf{w}^1_{a}  + Y_{2} \mathbf{w}^2_{a} , 
\end{equation*}
where $(\mathbf{w}^1_{a},\mathbf{w}^2_{a})$ is the coframe defined in \Eqsref{eq:referencecoframe}. Furthermore, since the $\mathbf{w}^1_{a}$ is the complex conjugated of $\mathbf{w}^2_{a}$, it follows that 
\begin{equation}\label{ec:relations_of_the_components}
Y_{2} = \mathbf{w}^a_2 \ Y_{a} = \overline{ \mathbf{w}^a_1 } \ Y_{a} = \overline{ \mathbf{w}^a_1  \ Y_{a} } = \overline{Y}_{1} .
\end{equation}
As a result, \Eqref{ec:prototipe0} can be further reduced to a $2\times2$ system of equations for the $2$-dimensional vector  $u = [X , Y_1]$ as follows
\begin{equation}\label{ec:prototipe}
\begin{split}
\partial_{r} u &= F( u, \eth u, \bar \eth u ,B ).\\
u|_{\rho_0} &= u_0,
\end{split}
\end{equation}
Given the spin weight properties of the frame $(\mathbf{e}^a_1, \mathbf{e}^a_2)$ and its coframe (refer to \Eqsref{eq:transformationrules_spin}), it becomes evident that $X$ and $Y_{1}$ possess spin weights of $0$ and $1$, respectively. Hence, we can express them in terms of the \textbf{SWSH} as follows:
\begin{eqnarray}\label{eq:functions_decomposed_inS2_ylm}
X(r,\theta,\phi) &=&  \sum\limits_{l,m}   \hspace{0.1cm}_{0} x_{lm}(r) \hspace{0.2cm}_{s}Y_{lm} (\theta,\varphi), \\
Y_1(r,\theta,\phi) &=& \sum\limits_{l,m} \hspace{0.1cm}_{1} y_{lm}(r) \hspace{0.2cm}_{s+1}Y_{lm} (\theta,\varphi).
\end{eqnarray}In a similar manner, it can be decomposed all the other tensorial components in $B$.  

To solve the system \Eqref{ec:prototipe} as an initial value problem numerically, Beyer et al. employed the fast spin-weighted spherical harmonic transform algorithm (\textbf{FswshT}) by Huffenberger and Wandelt \cite{huffenberger2010fast}. Their approach involved several steps: firstly, fixing a known $u$ at the initial surface $S_{\rho(r)}$. Secondly, using \textbf{FswshT} to determine the spectral coefficients of all tensor quantities within the equations. Thirdly, computing all terms involving eths operators through the relations in \Eqref{ec:eths_acting_on_components}. Finally, employing the inverse of the FswshT to compute the right-hand side of \Eqref{ec:prototipe}, facilitating the use of an explicit method (like a Runge-Kutta method) to approximate the solution $u$ for the subsequent radial step $\Delta r$, i.e., determining the functions $X$ and $Y_1$ at the hypersurface $S_{\rho(r+ \Delta r)}$.

However, a significant limitation in this implementation, as noted in \cite{beyer2017asymptotics}, is the inability to control the solution's asymptotic behavior or its boundary values near infinity. This limitation arises due to the usage of an implicit Runge-Kutta scheme to integrate the solutions in the direction of $\rho(r)$, allowing only an approximation up to a large value $\rho_f(r_f)$ and resulting in a lack of complete control over the solutions' asymptotic behavior. This issue is critical, particularly in obtaining asymptotically flat initial data sets $(h_{ab},\chi_{ab})$, widely regarded as essential for physically meaningful spacetimes (refer to Def. \ref{def:definition_asymptotically_flat}).

To exemplify this limitation, let's consider $\gamma_{\mu \nu},\alpha, \beta^\mu$, and $\mathring{K}_{\mu \nu}$ such that the metric components $h_{\mu \nu }$ satisfy \Eqref{eq:asymptotically_flat_condition1} in standard spherical coordinates. This condition implies that the topological spheres $S_{\rho}$ should become round as the radial coordinate $r$ tends to infinity. Now, assuming we solve \Eqref{ec:prototipe} as an initial value problem up to some sufficiently large $r_f$, utilizing $X(r)$, $Y_1(r)$, and \Eqref{ec:mean_curvature_decomposition}, we can retrieve the components $\chi_{\bar \mu \bar\nu }$ of the second fundamental form $\chi_{ab}$. However, due to the recursive determination of $X(r,\theta,\varphi)$ and $Y_1(r,\theta,\varphi)$ along the radial coordinate $r$, starting from some initial $r_0$ up to a finite value $r_f$, their asymptotic behavior remains uncontrollable. Consequently, the components $\chi_{\mu \nu }$ may or may not satisfy \Eqref{eq:asymptotically_flat_condition2}, posing uncertainty regarding the asymptotic flatness of the initial data.

Motivated by these limitations, in this work, we propose a modification by replacing the explicit Runge-Kutta scheme used for integrating $X(r,\theta,\varphi)$ and $Y_1(r,\theta,\varphi)$ along the $r$ coordinate with an infinite-element scheme. As detailed in the following section, this approach not only enables integration across the unbounded domain $[r_0, \infty)$ but also provides a means to control the asymptotic decay of the numerical solutions.

\section{A spectral-infinite element approach}\label{section:tercera_seccion}
This section aims to introduce our spectral-infinite element approach for solving the constraint equations in hyperbolic form. As this method involves a combination of two techniques, our presentation will be divided into two parts. Firstly, we will delve into the overarching concept of the infinite element method, a technique widely embraced by the scientific community for solving partial differential equations in unbounded domains (refer, for instance, to \cite{gerdes2000review}). Secondly, we will elucidate the amalgamation of this method with a spectral approach founded on the spin-weighted spherical harmonics basis. This combination forms the basis of our numerical infrastructure designed to address equations akin to \Eqref{ec:prototipe}.

\subsection{General idea of the infinite element method}\label{sec:inifiniteelementmethod}

Consider $y$ a square integrable real function defined on the unbounded domain $[r_0,\infty)$ for some  $r_0 \in \mathbb{R}^+$, i.e., $y \in L^2([r_0,\infty))$. Furthermore, let the bounded linear operator
\begin{equation*}
\mathcal{L}:= \dfrac{d }{dr} + f(r)
\end{equation*}
for some $f \in L^2([r_0,\infty))$. Then, assume that we want to find numerically $y$ such that  satisfies the differential equation
\begin{equation}\label{ec:31}
\mathcal{L} [ y(r) ] = g(r),
\end{equation}
with $g \in L^2([r_0,\infty))$ and constrained to the following two conditions
\begin{eqnarray}
y(r_0)  &=& y_0, \label{conditions12}\\  
y(r)  &=& \mathcal{O}({r}^{-1}) \label{conditions13},
\end{eqnarray}
with $y_0 \in \mathbb{R}$. The first condition represents a Dirichlet boundary condition, often straightforward to incorporate within the finite element method. Conversely, the second condition specifies a limitation on the solution's decay rate.
To enforce this decay condition numerically, we propose utilizing the infinite element method, a variant of the finite element method enabling the construction of numerical solutions featuring specific asymptotic decays. For a comprehensive understanding of this method, we recommend consulting \cite{zienkiewicz2000finite}.

We start by reformulating the problem into its variational form. Additionally, we will assume the existence and uniqueness of the solution $y$ within a linear space $V \subseteq L^2([a,\infty))$. Consequently, based on \Eqref{ec:31}, the subsequent equation should be valid for any test function $v \in V$:
\begin{equation}\label{ec:variationalform1}
\int_{r_0}^{\infty} \mathcal{L} [ y(r) ] v(r) \ dr = \int_{r_0}^{\infty} f(r) v(r) \ dr ,
\end{equation}
which is known in the literature as the variational form of the differential equation \Eqref{ec:31}. 

When applying the standard finite element method to numerically solve the variational problem, we encounter integrals defined over an unbounded interval. An initial approach to tackle this challenge involves truncating the interval at a sufficiently large value $b \in \mathbb{R}^+$ and solving \Eqref{ec:variationalform1} within the limited range $[r_0, b]$. However, ensuring an accurate solution necessitates the introduction of artificial boundary conditions at $b$ to preserve the asymptotic decay condition \Eqref{conditions13}.

The infinite element method presents itself as a modification of the standard finite element method tailored for handling variational problems in infinite domains. The method's fundamental concept involves utilizing a specific change of coordinates to transform the variational problem from an unbounded domain to a bounded domain, as elucidated below:

Consider the Zienkiewicz coordinate transformation (see for instance \cite{zienkiewicz1983novel})
\begin{equation}\label{ec:coordinate_transform_fem}
\xi = 1 - \dfrac{2 (r_0 - r_p)}{r - r_p}, \quad  r =  r_p  +\dfrac{2 (r_0 - r_p)}{1 -\xi},
\end{equation}
where $r_p$, the pole, is some real value $r_p \in (-\infty,r_0)$. An important property of this transformation is that it is not affected by changes of the origin of the coordinate system of $r$. For more details of its properties, see \cite{zienkiewicz2000finite}. Under \Eqref{ec:coordinate_transform_fem}, we have that the unbounded interval $[r_0,\infty)$ is   mapped to the bounded interval $[-1,1]$. Furthermore, if we discretize $[r_0,\infty)$ in $N$ points $r_i$, for $i=0,...,N-1$, it induces a discretization on $[-1,1]$ of $N+1$ points $\xi_i = \xi( r_i )$ where 
\begin{equation*}
\xi_0 = \xi( r_0 ) = -1 , \quad \xi_{N-1} = \xi(r_{N-1}) , \quad \xi_{N} = \xi( \infty) =1.
\end{equation*} 
Note that the  interval $[r_{N-1},\infty)$, which we will call as \textit{the infinite element}, is mapped to  $[\xi_{N-1},\xi_{N}]$. Replacing  \Eqref{ec:coordinate_transform_fem} into  \Eqref{ec:variationalform1},  
we obtain the following variational problem in the  bounded domain $[-1,1]$:
\begin{equation}\label{variationalequation_final}
\int_{-1}^{1} \mathcal{L} [ y(\xi) ] v(\xi) \ J(r;\xi) \ d\xi = \int_{-1}^{1} f(\xi) v(\xi) \ J(r;\xi) d\xi,
\end{equation}
where $J(r;\xi)$ denotes the Jacobian of the transformation, and conditions \Eqref{conditions12} and \Eqref{conditions13} imply  that,
\begin{eqnarray}
 y( 1)  &= y_0, \\  \label{boundaryinfinityconditions}
y(-1)  &= 0. \label{boundaryinfinityconditions2}
\end{eqnarray}
Consequently, due to the transformation to the new coordinate $\xi$, the variational problem is defined within a compact set featuring Dirichlet boundary conditions. This characteristic allows us to solve it using the standard finite element method in the following manner:

First, let us consider $\xi_0, ..., \xi_N$ be $N$-points on the interval $[-1,1]$ with $\xi_0=-1$ and $\xi_N=1$. Next, we choose a $N+1$-dimensional nodal bases $\{ \psi_i (\xi) \}_{i=0}^{N}$ of  polynomials compactly supported on $[-1,1]$ such that
\begin{equation}\label{ec:cspolynomials}
\psi_i(\xi_j) := \delta_{ij}, \quad  y(\xi) = \sum_{i=1}^{N} y(\xi_i) \psi_i(\xi). 
\end{equation}
Second, substituting the above in the variational form \Eqref{variationalequation_final} and taking the test functions $v(\xi)$ as $  \psi_j(\xi) $ for $j=0,...N$, we obtain the expression
\begin{equation*}
\sum_{i=1}^{N} \int_{-1}^{1}   \mathcal{L} [ y(\xi_i) \psi_i(\xi) ] \ \psi_j(\xi)  \ J(r;\xi) \ d\xi = \int_{-1}^{1} f(\xi) \   \psi_j(\xi) \ J(r;\xi) \ d\xi
\end{equation*}
Third, we write the above equation in the following matricial form 
\begin{equation}\label{ec:matricialfem}
A U = B, 
\end{equation}
where the components of the matrix $A$ and the vectors $b$ and $U$ are given by
\begin{eqnarray*}
A_{ji}&:=&  \int_{-1}^{1}  \psi_j(\xi)   \mathcal{L} [  \psi_i(\xi) ] \  J(r;\xi) \ d\xi , \label{ecs:Aij_fem}\\
B_{j} &:=& \int_{-1}^{1} f(\xi) \  \psi_j(\xi)  \ J(r;\xi) \ d\xi,  \label{ecs:Bj_fem}\\
U_{i} &:=& y(\xi_i).
\end{eqnarray*}
Please note that our selection of the test functions as $v = \psi_j(\xi)$ was arbitrary. However, this choice might require adjustment based on the specific problem and the properties of the resulting matrix $A$. For instance, ensuring that $A$ is a symmetric and positive definite matrix is desirable to guarantee the solvability of the system \Eqref{ec:matricialfem}. For further insights into this subject, refer to \cite{ern2013theory}.

Given that the polynomials $\psi_j(\xi)$ possess compact support within $[-1,1]$, they necessarily vanish outside this interval. However, within this range, they assume a general form, typically resembling $\xi$.
\begin{equation*}
\psi_j(\xi) =  a_0 + a_1 \xi + a_2 \xi^2 + ... \quad  \forall j = 0, ...,N. \quad  
\end{equation*}  
Hence, using the coordinate transformation \Eqref{ec:coordinate_transform_fem}, it follows that in the unbounded interval $[r_0,\infty)$ they get the form  
\begin{equation*}
\psi_i(r) =  b_0 + \dfrac{b_1}{r} + \dfrac{b_2}{r^2}  + ...,
\end{equation*} 
which clearly decay as $1/r$. Therefore, we obtain that the approximated solution $y(r)$ must have the form of
\begin{equation}
y(r) = \sum_{i=0}^{N-1} y(r_i) \psi_i(r) = c_0 + \dfrac{c_1}{r} + \dfrac{c_2}{r^2}  + ...  \ .
\end{equation} 
However, note that $c_0$ must be zero due to the boundary condition at infinity, i.e., $y(r) = \mathcal{O}({r}^{-1})$, as introduced by \Eqref{boundaryinfinityconditions2}. Moreover, the discrete points $r_i = r (\xi_i)$ obtained via \Eqref{ec:coordinate_transform_fem} are only meaningful for $i=0,...,N-1$, excluding the point at infinity.

To conclude this subsection, it is important to highlight that according to the general theory of the FEM (see, for example, \cite{brenner2007mathematical}), the approximation error of the numerical solution $y(\xi)$ across the entire domain should be $\mathcal{O}({h_{max}}^{p+1})$, where $p$ represents the order of the family of nodal polynomials $\psi_(\xi)$, and $h_{max}$ denotes the length of the largest element within $[-1,1]$.

\subsection{The spin-weighed spectral-infinite element method}

In this section, we aim to extend the infinite element method to differential equations resembling \Eqsref{ec:prototipe}, which are defined over the unbounded domain $[r_0,\infty) \times \mathbb{S}^2$. To achieve this, we will employ the spin-weighted spherical harmonics basis. This approach allows us to reduce a discrete three-dimensional problem into a set of one-dimensional problems, which can be addressed using the infinite element method detailed in the previous section. Our presentation will commence by introducing the central concept of the method: the spin-weighted spectral decomposition of discrete square-integrable functions defined on $[r_0,\infty) \times \mathbb{S}^2$. Subsequently, we will utilize this concept to develop an algorithm for solving differential equations resembling \Eqsref{ec:prototipe}, encompassing both the linear and non-linear scenarios.

\subsubsection{Spectral decomposition of discrete functions }\label{sec:subsectionspectraldecomposition}

Let us consider standard spherical coordinates  $(r,\theta,\phi)$ to parameterize the domain $[r_0,\infty) \times \mathbb{S}^2$, with some positive value $r_0$. This domain is discretized by taking   $N_r \times N_\theta \times N_\phi$ points of the form ${(r_i,\theta_j,\phi_k)}$ for $i=1,...,N_r$, $j=1,...,N_\theta$ and $k=1,...,N_\phi$, respectively. Furthermore, we choose   $\theta_j,\phi_k$  equally spaced as: 
\begin{equation*}
\theta_j = \dfrac{ \pi j}{N_{\theta}}, \quad \phi_k = \dfrac{ 2 \pi k}{N_{\phi}}.
\end{equation*}  
On the other hand, the points $r_i$ are chosen arbitrary (clearly we omit infinity). From now on, we will  refer to this set of points $(r_i,\theta_j,\phi_k)$ as the \textit{mesh-points}. Next, consider  a function $f \in L^2([r_0,\infty)\times \mathbb{S}^2)$ sampled on the mesh points. Then, from \Eqref{ec:spectral_decomposition} it follows that
\begin{equation*}\label{eq:functionS2_discrete}
f(r_i,\theta_j,\phi_k) =  \sum\limits_{l,m}  \hspace{0.1cm}_{s}a_{lm}(r_i) \hspace{0.2cm}_{s}Y_{lm} (\theta_j,\varphi_k),
\end{equation*} 
where the sum over the index $l$ is truncated at some  positive interger $L$, known as the \textit{band-limited},  such that $N_\theta, N_\phi \geq 2 L + 1$ (see \cite{huffenberger2010fast} for details). The ${}_{s}a_{lm}(r_i)$ are the spectral coefficients along the discrete radial points $r_i$.   Note that we can find them at any fixed $r_i$ by applying the \textbf{FswshT} over $f$ sampled at the sphere of radius $r_i$. Therefore, if we know  the function $f$  at the mesh points, we can also find the spectral coefficients $a_{lm}(r_i)$ in the discrete radial points. Moreover, since this transform (and its inverse) is based on the computation of two Fourier transforms (see  \cite{huffenberger2010fast}), we can obtain all the ${}_{s}a_{lm}(r_i)$ values from $f$, sampled at the mesh-points, by computing approximately $(N_{\theta}\text{log}_2 N_{\theta}) \times (N_{\phi} \text{log}_2 N_{\phi}) \times N_{r}$ operations.

\subsubsection{The linear case}\label{sec:linear_case_algorithm}
In this section, we present an algorithm designed to handle equations akin to \Eqref{ec:prototipe}, specifically when it can be reduced to a linear equation.

To begin with, suppose we seek a function $y\in L^2([r_0,\infty)\times \mathbb{S}^2)$ that approximate the solution of the following linear PDE:
\begin{equation}\label{ec:general_linear_eq}
\partial_{r} y   =  f \ \eth y + g  \ \bar{\eth} y  + h \  y  + q ,
\end{equation}
for some given functions $f,g,h,q \ \in L^2([r_0,\infty)\times \mathbb{S}^2)$ and constrained to the conditions
\begin{equation}\label{ec:linear_case_BC} 
\begin{split}
y(  r_0  )  &= y_0, \\  
y( r )  &= \mathcal{O}({r}^{-1}).
\end{split}
\end{equation}
Using the spin-weighted spherical harmonic decomposition \Eqref{ec:spectral_decomposition} for functions in $L^2([r_0,\infty)\times \mathbb{S}^2)$, and the properties of the eth-operators discussed in \Sectionref{sec:swsh_seccion1}, we have that 
\begin{eqnarray}\label{eq:functionS2}
y &=&  \sum\limits_{l_1,m_1}  {}_{s}a_{l_1 m_1}(r) \hspace{0.2cm}_{s}Y_{l_1 m_1} (\theta,\varphi), \label{ec:forthesolution}\\
\partial_r y &=&  \sum\limits_{l_1,m_1} \partial_r(  {}_{s}a_{l_1 m_1}(r) ) \hspace{0.2cm}_{s}Y_{l_1 m_1} (\theta,\varphi),  \\
\eth y &=&  \sum\limits_{l_1 , m_1} S(l_1,s, 1) \
 {}_{s}a_{l_1 m_1}(r) \hspace{0.2cm}_{s+1}Y_{l_1 m_1} (\theta,\varphi),\\
\bar \eth y &=&  \sum\limits_{l_1 , m_1}  S(l_1,s, -1) \
 {}_{s}a_{l_1 m_1}(r) \hspace{0.2cm}_{s-1}Y_{l_1 m_1} (\theta,\varphi). 
\end{eqnarray}
As it was mentioned in \Sectionref{sec:swsh_seccion1}, when we multiply two functions of certain spin weight, say $s_1$ and $s_2$, the resulting function must have spin-weight $s_1+s_2$. Therefore,  to balance the spin-weighted in both sides of the equation, the functions $f,g,h \in L^2([r_0,\infty)\times \mathbb{S}^2)$ should be decomposed as
\begin{eqnarray}
f &=&  \sum\limits_{l_3,m_3}  {}_{-1}b_{l_2 m_2}(r) \hspace{0.2cm}_{-1}Y_{l_2 m_2} (\theta,\varphi),\\
g &=&  \sum\limits_{l_3,m_3}  {}_{1}c_{l_3 m_3}(r) \hspace{0.2cm}_{ 1}Y_{l_3 m_3} (\theta,\varphi),\\
h &=&  \sum\limits_{l_2,m_2} {}_{0}d_{l_4 m_4}(r) \hspace{0.2cm}_{0}Y_{l_4 m_4} (\theta,\varphi), \\
q &=&  \sum\limits_{l_4,m_4} {}_{s}e_{l_5 m_5}(r) \hspace{0.2cm}_{s}Y_{l_5 m_5} (\theta,\varphi). \label{ec:forthesolution2}
\end{eqnarray}
Using the inner product \Eqref{integral_properties_spherical_harmonics}, we project the equation \Eqref{ec:general_linear_eq} to the basis $\ _{s} Y_{l m}(\theta,\varphi)$
as follows
\begin{equation}\label{equation_galerkind}
\begin{split}
\langle \  \partial_{\rho} y \ , \ _{s} Y_{l m}(\theta,\varphi) \ \rangle
   &=  \langle \ f \ \eth y \  , \ _{s} Y_{l m}(\theta,\varphi) \ \rangle + 
    \langle \ g \ \bar \eth y \   , \ _{s} Y_{l m}(\theta,\varphi) \ \rangle \\
   &  + \langle \ h \ y  , \ _{s} Y_{l m}(\theta,\varphi) \ \rangle + \langle \ q \ , \ _{s} Y_{l m}(\theta,\varphi) \ \rangle 
\end{split}   
\end{equation}
for $0 \le l \le L$, $-l \le m \le l$ and some fixed $s$, which corresponds to the spin weight of the unknown function $y$. Below, we examine these five inner products by separate.

Because of \Eqsref{ec:forthesolution} - (\ref{ec:forthesolution2}), and the orthogonality of the SWSH \Eqref{integral_properties_spherical_harmonics}, it can be obtained that
\begin{equation}\label{ec:galerking1}
\langle \  \partial_{r} y \ , \ _{s} Y_{l m}(\theta,\varphi) \ \rangle =
\partial_{\rho} ( {}_{s}a_{l  m }(r) )
\end{equation}
and
\begin{equation}\label{ec:galerking2}
\langle \ q \ , \ _{s} Y_{l m}(\theta,\varphi) \ \rangle =   {}_{s}e_{l m }(r).
\end{equation}
Furthermore, it is clear that 

\begin{equation}
\begin{split}
& \langle \ f \ \eth  y \ , \ _{s} Y_{l m}(\theta,\varphi) \ \rangle  = 
\int \limits_{\mathbb{S}^2}   f \ \eth  y  \   {}_{s} \overline{Y}_{l m }(\theta,\varphi) \s d\Omega =  \\
& \sum\limits_{l_1 , m_1 }    \sum\limits_{l_2 , m_2 }  S(l_1,s, 1) \ {}_{s}a_{l_1 m_1}(r)
\  {}_{-1}b_{l_2 m_2}(r) 
\int \limits_{\mathbb{S}^2}
{ }_{-1}Y_{l_2 m_2} (\theta,\varphi) \
{ }_{s+1}Y_{l_1 m_1} (\theta,\varphi) \
{ }_{s} \overline{Y}_{l m }(\theta,\varphi) \s d\Omega \label{equacion_de_paso} .
\end{split}
\end{equation}

To reduce the above expression, we introduce the $\mathcal{C}$-functions $\mathcal{C}_{s_i l_i m_i, s l m}$ that act over sets of spectral coefficients $\{ {}_{s_j} z_{l_j m_j}(r) \}$ as
\begin{equation}\label{ec:C-equation}
\begin{split}
\mathcal{C}_{s_i l_i m_i, s l m} ( \ \{  {}_{s_j}z_{l_j m_j}(r) \} \ ) =&  \sum\limits_{l_j , m_j } {}_{s_j} z_{l_j m_j}(r) \ (-1)^{s+m} \
\sqrt{ \dfrac{ (2 l_i +1)(2 l_j +1)(2 l  +1) }{4 \pi} } \\
   & \begin{pmatrix}
   l_i & l_j & l\\
   m_i & m_j & -m
   \end{pmatrix} \begin{pmatrix}
   l_i & l_j  & l\\
   s_i & s_j &  -s
   \end{pmatrix}. 
\end{split}
\end{equation}

Then,  replacing \Eqsref{ec:clebsh_jordan_formula} and (\ref{ec:complex_conjugate_swsh}) into the right hand side of \Eqref{equacion_de_paso} we obtain
\begin{equation}\label{ec:galerking3}
\langle \ f \ \eth  y \ , \ _{s} Y_{l m}(\theta,\varphi) \ \rangle = \sum\limits_{l_1 , m_1 } \mathcal{C}_{(s+1) l_1 m_1,s l m} ( \ \{  {}_{-1}b_{l_2 m_2}(r) \} \ )  \ S(l_1,s, 1) \ {}_{s}a_{l_1 m_1}(r).
\end{equation}
Similarly, we can write the rest of the products  as
\begin{eqnarray}
\langle \ g \ \bar \eth  y \ , \ _{s} Y_{l m}(\theta,\varphi) \ \rangle & = & \sum\limits_{l_1 , m_1 }    
   \mathcal{C}_{(s-1) l_1 m_1,s l m} ( \ \{  {}_{1}c_{l_3 m_3}(r) \} \ ) \ S(l_1,s, -1) \  {}_{s}a_{l_1 m_1}(r), \label{ec:galerking4} \\
\langle \ h \    y \ , \ _{s} Y_{l m}(\theta,\varphi) \ \rangle & = & \sum\limits_{l_1 , m_1 }    
 \mathcal{C}_{(s) l_1 m_1,s l m} ( \ \{  {}_{0}d_{l_4 m_4}(r) \} \ )   \   {}_{s}a_{l_1 m_1}(r)  \label{ec:galerking5}.
\end{eqnarray}  
Note that the  $\mathcal{C}$-functions account for the coupling between the spectral coefficients. 

Finally, substituting \Eqsref{ec:galerking1}, (\ref{ec:galerking1}), (\ref{ec:galerking2}), (\ref{ec:galerking3}), (\ref{ec:galerking4}) and  (\ref{ec:galerking5}) into \Eqref{equation_galerkind},  we obtain a set of  one-dimensional coupled differential equations for the spectral coefficients ${ }_{s} a_{l, m}(r)$:
\begin{equation}\label{ec:main_spectral_equation}
\begin{split}
\partial_{r} ( { }_{s }a_{l  m }(r) ) &= \sum\limits_{l_1 , m_1 }  
 \Bigg(  \mathcal{C}_{(s+1) l_1 m_1,s l m} ( \ \{  {}_{-1}b_{l_2 m_2}(r) \} \ )  \ S(l_1,s, 1)  \\ 
 &  \qquad +  \mathcal{C}_{(s-1) l_1 m_1,s l m} ( \ \{  {}_{1}c_{l_3 m_3}(r) \} \ ) \ S(l_1,s, -1)  \\
 &  \qquad +  \mathcal{C}_{(s) l_1 m_1,s l m} ( \ \{  {}_{0}d_{l_4 m_4}(r) \} \ ) \
   \Bigg) \ { }_{s} a_{l_1 m_1}(r) + { }_{s} e_{l m}(r).  
\end{split}
\end{equation} 
Because of $ -l \leq m \leq l $ for all $l$, the total number of equations is  $\sum\limits_{l=|s|}^{L} (2l+1)$. Note that the boundary conditions for each spectral coefficient ${ }_{s} a_{l_1 m_1}(r)$ are  obtained from \Eqref{ec:linear_case_BC} as 
\begin{equation*}
\begin{split}
{}_{s}a_{l m}(r_0) &= {}_{s} \bar a_{l m},  \\
{}_{s}a_{l m}(r) &= \mathcal{O}({r}^{-1}),
\end{split}
\end{equation*}
where the $\bar a_{l m}$  are the spectral coefficients of the initial data $y_0$. \Eqsref{ec:main_spectral_equation} comprise a coupled system of linear equations with the form of \Eqref{ec:31}, hence, it can be solved numerically on the mesh-points by employing the infinite element method explained in \Sectionref{sec:inifiniteelementmethod}. For doing that, however, we have to know the values of the spectral coefficients at the radial discrete point of the domain, i.e., we need to know the spectral decomposition of the functions $f,g,h,q$ on the mesh-points as it was explained in the previous  \Sectionref{sec:subsectionspectraldecomposition}.

To summarize the above, we can  solve a linear PDE with the form of \Eqref{ec:general_linear_eq} by following the next algorithm: 
\begin{enumerate}[(i)]
\item Define the mesh $(r_i,\theta_j,\phi_k)$ on $[r_0,\infty) \times \mathbb{S}^2$ as   explained in \Sectionref{sec:subsectionspectraldecomposition}. 
\item Compute 
the spectral decomposition of the functions $f,g,h,q$ on the mesh-points $(r_i,\theta_j,\phi_k)$.
\item Using the $\mathcal{C}$-functions defined on \Eqref{ec:C-equation}, compute the coefficients of the system \Eqsref{ec:main_spectral_equation}  for each value of $r_i$.
\item Use the infinite element method explained in \Sectionref{sec:inifiniteelementmethod} for solving numerically the coupled linear system of equations.
\item Finally, using the inverse SWSH transform (or directly \Eqref{ec:forthesolution}),  recover the solution $y(r)$ for all the mesh-points $(r_i,\theta_j,\phi_k)$. 
\end{enumerate}

\subsubsection{The non-linear case}
It is widely recognized that both the Fixed-point method and the Newton-Raphson method are standard approaches for solving non-linear PDEs. However, these methods rely on approximating the solutions of the non-linear PDE by iteratively solving other linear PDEs. Therefore, in numerical terms, solving a non-linear PDE using one of these methods involves solving several instances of a linear PDE. For an in-depth exploration of these methods applied to the solution of PDEs, one can refer to \cite{gockenbach2006understanding}.

In this subsection, we will specifically discuss how to utilize the algorithm introduced for solving linear PDEs in the preceding sub-section to numerically solve non-linear PDEs such as:
\begin{equation}\label{ec:no-linear_ec}
\partial_{\rho} y = F(y, \eth y,\bar \eth y),  
\end{equation}
constrained to the restrictions
\begin{eqnarray*} 
y( 1 ) &=& y_0, \\  
y( r ) &=& \mathcal{O}({r}^{-1}),
\end{eqnarray*}
by both the Fixed-point method and the Newton-Raphson method.\\

\textbf{Fixed-point method}: 
Assuming that the function $F(y, \eth y,\bar \eth y)$ is Lipschitz under $y$ (see  \cite{stoer2013introduction}), it can be proved that the sequence of $\{ y^{(n)} \}$ defined by
\begin{equation}\label{fixed_point_equations}
\partial_{\rho} y^{(n+1)} = F(y^{(n)},\eth y^{(n)},\bar \eth y^{(n)}), 
\end{equation}
converges to the solution of \Eqref{ec:no-linear_ec} for some given $y^{(0)}$. Note that we can solve the above equation implementing the algorithm given in \Sectionref{sec:linear_case_algorithm} because it takes the form of a linear PDE of the form \Eqref{ec:general_linear_eq} with $f,g,h$ equal to zero and $q=F(y^{(n)},\eth y^{(n)},\bar \eth y^{(n)}) $, i.e.,
\begin{equation*}
\partial_{\rho} y^{(n+1)} = q.
\end{equation*}
It is important to note that in this scenario, there is no need to compute the $\mathcal{C}$-functions, resulting in a significant simplification of the algorithm. With this approach, the numerical solution to \Eqref{ec:no-linear_ec} involves an iterative solution of \Eqref{fixed_point_equations}. The iteration proceeds until the difference between the solutions $y^{(n+1)}$ and $y^{(n)}$ (measured under a certain norm) becomes less than or equal to a predefined tolerance.

Finally, we have to mention that even though this method is relatively easy to implement, its convergence rate is only of linear order (see for instance \cite{wriggers2008nonlinear}). \\

\textbf{Newton-Raphson method}: This method is essentially the fixed-point with some modifications. The idea consists of assuming that the approximations $y^{(n+1)}$ are given by 
\begin{equation}\label{update_solution}
y^{(n+1)} = y^{(n)} +  \delta y,
\end{equation}
where  $y^{(n)}$ is a known approximation and $\delta y$ is a small quantity that is determined  by solving the linearization of  \Eqref{ec:no-linear_ec} 
 \begin{eqnarray}\label{newton-equations}
\partial_{r} \delta y  &=&  \dfrac{\partial \    }{\partial \ \eth y} \left[ F(y, \eth y,\bar \eth y) \right] \bigg|_{y=y^{(n)}}   \eth \delta y +  
 \dfrac{\partial \   }{\partial \ \bar\eth y} \left[ F(y, \eth y,\bar \eth y) \right] \bigg|_{y=y^{(n)}}   \bar\eth \delta y  \nonumber \\
   & & + 
 \dfrac{\partial \   }{\partial \ y} \left[ F(y, \eth y,\bar \eth y) \right] \bigg|_{y=y^{(n)}}    \delta y . \label{ec:linearization_for_newton}
\end{eqnarray}
It is worth noting that this equation is akin to the linear PDE \Eqref{ec:general_linear_eq} with $q=0$, allowing us to apply the algorithm described in \Sectionref{sec:linear_case_algorithm} for its solution.

Similar to the fixed-point method, we solve \Eqref{newton-equations} iteratively (updating the approximations via \Eqref{update_solution}) $n$ times until the discrepancy between the approximations $y^{(n+1)}$ and $y^{(n)}$ falls below a specified tolerance. The notable advantage of this method over the fixed-point method is its faster convergence rate—it converges quadratically when the initial guess $y^{(0)}$ is sufficiently close to the exact solution. However, its implementation is more involved due to the need for calculating \Eqref{ec:linearization_for_newton} and computing the $\mathcal{C}$-functions (for more details, refer to texts such as \cite{stoer2013introduction}).

\section{Application of the spectral-infinite element method}\label{section:cuarta_seccion} 
In this section, we aim to utilize the spectral method outlined in the previous section to calculate perturbations of a rotating black hole with small angular momentum, ensuring asymptotic flatness. We'll commence by revisiting the Kerr metric expressed in Kerr-Schild coordinates, emphasizing its simplification when dealing with small angular momentum scenarios. Subsequently, we'll employ this simplified metric as a background to linearize the system \Eqref{ec:prototipe} and showcase several numerical solutions.

\subsection{The Kerr black hole with small angular momentum}\label{subsection:smallangularmomentun}
We will delve into the Kerr metric expressed in what's known as the Kerr-Schild form. This choice of metric representation is motivated by the ability to identify non-time symmetric Cauchy surfaces. Specifically, it results in the second fundamental form $\chi_{ab}$ of the Cauchy surfaces $\Sigma$ relative to the four-dimensional spacetime $M$ not equating to zero. This condition is vital for the hyperbolic formulation of the constraints, as discussed in \Sectionref{sec:decompositionconstraints} via \Eqsref{eq:Constraint3}, (\ref{eq:Constraint4}), and (\ref{eq:Constraint5}).

Let is examine the manifold $(M,g_{ab})$, which represents the Kerr spacetime. Employing the adapted Kerr-Schild coordinates $(t,x,y,z)$, the Kerr metric is typically expressed as (for a comprehensive treatment, refer to \cite{baumgarte2010numerical}):

\begin{equation}\label{ec:kerrblackholenull}
  g_{ab} = \eta_{ab} + 2 H  l_{a} \otimes l_{b} , 
\end{equation}
where $ H $ is  given by
\begin{equation*}\label{def:H}
H =  \dfrac{r M}{r^2 + a^2 z^2 / 2}, 
\end{equation*}
and $r$ is a function depending on coordinates $(x,y,z)$ that satisfies the relation 
\begin{equation*}\label{eq:relation_r_and_coords}
  \dfrac{ x^2 + y^2 }{ r^2 + a^2 } + \dfrac{z^2}{r^2} = 1 .  
\end{equation*}
On the other hand, $l_a$  is a null covector with respect to both $g_{a b}$ and the Minkowski metric $\eta_{ab}$ which has components with respect to the coordinate frame $(\partial^a_t,\partial^a_x,\partial^a_y,\partial^a_z)$ as
\begin{equation*}\label{def:li}
  l_{\hat\mu} = \left( 1, \dfrac{r x + a y} { r^2 + a^2} ,  \dfrac{r y - a x }{ r^2 + a^2} , \dfrac{ z }{ r }  \right),
\end{equation*}
where we have used $\hat\mu,\hat\nu,...$ to denote spacetime indices. The relation  between the spatial Kerr-Schild coordinates $(x,y,z)$ and the standard spherical coordinates $(r,\theta,\varphi)$ are given by 
\begin{equation*}\label{coordinate_transformations}
  x = (r \cos \phi + a \sin \phi ) \sin \theta, \quad 
  y = (r \sin \phi - a \cos \phi ) \sin \theta, \quad
  z = r \cos \theta . 
\end{equation*}
Replacing these transformations into \Eqref{ec:kerrblackholenull}, one can easily obtain the Kerr metric in terms of the coframe $(\df t_a,\df r_a,\df \theta_a,\df \varphi_a)$ as
\begin{eqnarray*}\label{Kerr_metric}
g_{ab} = \left( -1 + \frac{2 M r}{P^2} \right)  \df t_a  \otimes \df t_b + \frac{4 M r}{P^2}  \df t_a \otimes \df r_b  -  \left( \frac{4 a M r \sin ^2 \theta }{P^2} \right)  \df t_a   \otimes \df \varphi_b  \nonumber\\ 
 +  \left( 1 +\frac{2 M r}{P^2} \right)\df r_a \otimes \df r_b 
  -  \left( 2\frac{a \sin ^2 \theta \left(2 M r+P^2\right)}{P^2} \right) \df r_a \otimes \df \varphi_b \nonumber\\ 
 +  P^2 \df \theta_a  \otimes \df \theta_b
 + \left( \frac{\sin ^2 \theta  \left(\left(a^2+r^2\right)^2-a^2 \Delta  \sin ^2 \theta \right)}{P^2} \right)  \df \varphi_a \otimes \df \varphi_b,
\end{eqnarray*}
with $\Delta = a^2-2 M r+r^2$,   $P = \sqrt{a^2 \cos ^2 \theta +r^2}$, $M$ is a positive quantity that represents the black hole mass and $a$ its angular momentum. Note that $\df t_a = \nabla_a t$, $\df x_a = \nabla_a x$, $\df y_a = \nabla_a y$, and $\df z_a = \nabla_a z$, where $\nabla_a$ represents the covariant derivative compatible with the Kerr metric.

It is commonly known that the standard Schwarzschild metric is obtained by setting $a = 0$ (refer to \cite{Wald:1984un}). In the context of a rotating black hole with small angular momentum, we assume $a^2 \approx 0$. Under this approximation, the Kerr metric takes the reduced form:
\begin{equation}\label{Kerr_metric0}
\begin{split}
g_{ab} =& \left( \frac{2 M}{r} - 1 \right) \df t_a  \otimes \df t_b+ \frac{4 M  }{r}  \df t_a \otimes \df r_b  -  \left( \frac{4 a M \sin ^2 \theta }{r} \right)  \df t_a  \otimes \df \varphi_b   +  \left( 1 +\frac{2 M }{r} \right)\df r_a \otimes \df r_b \\
& -  \left( 2\frac{a \sin ^2 \theta \left(2 M + r\right)}{r} \right) \df r_a \otimes \df \varphi_b +  r^2 \df \theta_a \otimes \df \theta_b + r^2 \sin^2\theta \ \df \varphi_a \otimes \df \varphi_b.
\end{split}
\end{equation}
From now on, we will refer to this metric as  \textit{the Kerr metric with small angular momentum} (\textbf{ksa}).

\subsection{The background solutions}
Utilizing the standard $3+1$ decomposition over the spacetime $(M,g_{ab})$ (as discussed in detail in \cite{baumgarte2010numerical}), we can derive from \Eqref{Kerr_metric0} the expressions for the induced metric $h_{ab}$ and the second fundamental form $\chi_{ab}$ of the $t$-constant Cauchy surfaces $\Sigma$ within $M$, which are given by:
\begin{eqnarray}\label{Kerr_metric_cauchy}
  h_{ab} &=&   \left( 1 +\frac{2 M }{r} \right)\df r_a \otimes  \df r_b 
 -  \left( 2 \frac{ a \sin ^2 \theta \left(2 M + r\right)}{r} \right) \df r_a \otimes \df \varphi_b  + r^2 \mathring \gamma_{ab}, \\
  \chi_{ab}&=& 2 M  \sqrt{ 1 + \dfrac{2 M}{r} } \left( \dfrac{ M + r}{r^2 (2 M + r)} \df r_a \otimes  \df r_b - \dfrac{a \sin^2 \theta}{r^2} \ \df r_a \otimes \df \varphi_b  
- \dfrac{r}{  r + 2 M  }   \mathring \gamma_{ab}  \right)  ,
\end{eqnarray}
where $\mathring \gamma_{ab}$ is the standard metric of the round two-sphere in spherical coordinates 
\begin{equation*}
\mathring \gamma_{ab} = \df \theta_a \otimes \df  \theta _b +  \sin ^2\theta  \df \varphi_a \otimes \df \varphi_b. 
\end{equation*}
In the subsequent analysis, we proceed with the $2+1$ decomposition, detailed in \Sectionref{sec:TheSytemOfEquations}, to foliate the Cauchy surfaces $(\Sigma, h_{ab})$ into topological spheres $S_{\rho}$ for determining suitable tensors $\gamma_{ab},\alpha, \beta^{a}$, and $\mathring{K}_{ab}$. These tensors will facilitate the computation of exact solutions $X$ and $Y_a$ for the hyperbolic system \Eqref{ec:prototipe}. Consequently, these solutions will provide the background values of $X$ and $Y_a$ that satisfy the hyperbolic constraint equations, yielding initial data for the Kerr spacetime with small angular momentum.

We initiate the process by conducting the $2+1$ decomposition on $(\Sigma, h_{ab})$, as elucidated in \Sectionref{sec:TheSytemOfEquations}. Subsequently, for expressing all tensor components in terms of the \textbf{swsh} and eth-operators, we select the foliations $S_{\rho}$ and non-coordinate frames detailed in \Sectionref{sec:pseudo_spectral_method}. Through a direct computation, we can represent the pair $(h_{ab},\chi_{ab})$ in terms of the non-coordinate coframe $(\mathbf{w}_0,\mathbf{w}_1,\mathbf{w}_2)$, where $\mathbf{w}^1_a$ and $\mathbf{w}^2_a$ denote the coframe vectors defined in \Eqref{eq:referencecoframe}, and $\mathbf{w}^0_a := \df r_a$. The expressions for these quantities are given by:
\begin{eqnarray}\label{Kerr_metri}
  h_{ab} &=& \left( 1 +\frac{2 M }{r} \right) \mathbf{w}^0_a \otimes \mathbf{w}^0_b 
+ \left( 2 \dfrac{\text{i} a ( 2M+r) \sin \theta }{ \sqrt{2} r  }  \right) \mathbf{w}^0_a \otimes  ( \mathbf{w}^1_b -  \mathbf{w}^2_b ) + r^2 \mathring \gamma_{ab}, \\ \label{ec:firstfundamentalformsp}
\chi_{ab}&=& 2 M  \sqrt{ 1 + \dfrac{2 M}{r} } \Bigg( \dfrac{M+r}{r^2(2M+r)} \mathbf{w}^0_a \otimes \mathbf{w}^0_b + \left( \dfrac{ \text{i} a \sin \theta}{ \sqrt{2} r^2} \right) \mathbf{w}^0_a \otimes  ( \mathbf{w}^1_b -  \mathbf{w}^2_b ) + r^2 \mathring \gamma_{ab}  \Bigg) ,\label{ec:second_fundamental_form_to_be_used} \label{ec:secondfundamentalformsp}
\end{eqnarray}
where $\mathring \gamma_{ab} = 2 \mathbf{w}^1_a \otimes  \mathbf{w}^2_b$ is the metric of the round two-sphere. Next, we choose the lapse and the shift vector as  \begin{equation}\label{alpfaybetaforkerr}
\alpha = \sqrt{1+ \dfrac{2M}{r} },  \quad \text{and} \quad \beta^{a} = \dfrac{-\text{i} a ( 2M+r) \sin \theta }{ \sqrt{2} r^3 } \left( \mathbf{e}^a_1 - \mathbf{e}^a_{2}  \right),
\end{equation}
from where we obtain that the normal vector to hypersurfaces $S_{\rho}$ and its corresponding  covector are, respectively (see \Eqref{eqc:1}),
\begin{equation}\label{ec:normalVector_frame}
n^a = \alpha^{-1} \left( \mathbf{e}^a_0  - \beta^a \right),  \quad n_{a}= \alpha \mathbf{w}^0_a.
\end{equation}
Further, from \Eqref{ec:firstfundamentalformsp} and using \Eqsref{eqc:2} and (\ref{ec:segundaformaenspheres}), we obtain that the induced metric $\gamma_{ab}$ and the second fundamental form $H_{ab}$ of the foliations $S_{\rho}$ take the form
\begin{equation}\label{kecuationforthemetricinerr}
\gamma_{ab} = r^2 \mathring \gamma_{ab}, \quad H_{ab} = \dfrac{r}{\alpha}  \mathring \gamma_{ab}.
\end{equation}
Consequently, we establish that the hypersurfaces $S_{\rho}$ manifest as round two-spheres with a radius of $r$. Consequently, $\rho(r) = r$, and thereby, $\partial^a_{\rho} = \partial^a_r$. Pursuing the decomposition of the tensor $\chi_{ab}$ as illustrated in \Eqref{ec:mean_curvature_decomposition}, we derive from \Eqref{ec:second_fundamental_form_to_be_used} the following expression:
\begin{equation}\label{ec:importanteqcuations}
 X^{(K)} = \dfrac{-4M}{\alpha r^2} , \quad  Y^{(K)}_a = \dfrac{ \text{i}  a 3  M \sin \theta}{\sqrt{2} r^2} \left( \mathbf{w}^1_a - \mathbf{w}^2_a \right), \quad \mathring K_{ab}=0.
\end{equation}
According to \Eqref{eq:Constraint3}, the quantity $Z$ is determined by $X$, $Y_{a}$ and the function $\kappa_{0}$, which has the form of
\begin{equation}\label{equation_for_the_kappa}
\kappa_{0} = \dfrac{ 8 M^2 }{r^2 (2 M + r)^2 }.
\end{equation}
Note that  by construction $X^{(K)}$ and $Y^{(K)}_{\mu}$  are the solutions of  \Eqsref{eq:Constraint4} and (\ref{eq:Constraint5}) respectively. Therefore, the vector $ u^{(K)}:=(X^{(K)}  , Y^{(K)}_{\mu})$ is an exact solution of the system  \Eqref{ec:prototipe}. From now on, we will refer to $u^{(K)}$ as \textit{the background solutions}.

\subsection{The system for small perturbations}

Because of the conformal factor between the induce metric $\gamma_{ab}$ and the standard round two-sphere is $r^2$, it follows that all the conformal coefficients $C^{\lambda}_{\nu\mu}$ defined in \Eqref{ec:conformalcoefficients} vanish. Thus, from  \Eqref{ec:relation_covariant_derivative_eth} the components of tensors $D_a T^{b...c}_{\quad \ \ d...f}$ take the following form:
 \begin{equation} 
\begin{split}
D_{\tau} T^{\mu...\nu}_{\qquad  \sigma...\lambda} &= \dfrac{1}{\sqrt{2}} \eth_\tau T^{\mu...\nu}_{\qquad  \sigma...\lambda},
\end{split}
\end{equation}
where the terms $\eth_{\tau}$ are determined by \Eqref{ec:indiceseths}. In other words, the covariant derivatives on the hypersurfaces $S_{\rho}$ are completely determined by the eth-operators.  Using this fact, and substituting the tensor components of $\alpha,\beta^{a}, \gamma_{ab}, \mathring{K}_{ab}$  given in \Eqsref{alpfaybetaforkerr}, (\ref{kecuationforthemetricinerr}) and (\ref{ec:importanteqcuations}) into the tensorial equations \Eqsref{eq:Constraint4} and (\ref{eq:Constraint5}) we obtain  

\begin{equation}\label{eq:Constraint44}
\mathscr{L}_{n} X - \gamma^{\mu\nu} \dfrac{1}{\sqrt{2}} \eth_\nu Y_\mu   - \left( Z - \dfrac{1}{2} X \right) H =0,  
\end{equation}
\begin{equation}\label{eq:Constraint55}
\mathscr{L}_{n} Y_{\mu}  + \dfrac{1}{ X \sqrt{2} } \left( Z \eth_\mu X - 2  Y^{\nu} \eth_{\mu} Y_{\nu} \right) 
+ H Y_{\mu}  = 0,    
\end{equation}   
where 
\begin{equation}
Z =  \dfrac{Y^{\mu} Y_{\mu}}{2 X}  - \dfrac{X}{4} -  \dfrac{\kappa_0 }{2X}  , \quad H = \dfrac{2}{\alpha r},
\end{equation}
and the Lie derivatives computed in terms of the covariant derivatives are given by
\begin{equation}\label{eq:casi_linear1}
\mathscr{L}_{n} X = \dfrac{1}{ \alpha \sqrt{2}}  (\sqrt{2} \ \partial_{r} X  -  \beta^{ \nu} \eth_{ \nu} X ),
\end{equation}
\begin{equation}\label{eq:casi_linear2}
\mathscr{L}_{n} Y_{ \mu} = \dfrac{1}{ \alpha \sqrt{2}} (\sqrt{2} \ \partial_{r} Y_{ \mu} -   \beta^{  \nu } \eth_{  \nu } Y_{ \mu} -    Y_{
 \nu} \eth_{ \mu}     \beta^{  \nu } ).
\end{equation}
Note that we have used the fact that $\dot{n}^\mu = -D_{\mu}(\ln\alpha)$ and $D_{\mu}R$ vanish because $\alpha$ and $R$ only depend on the $r$ coordinate. Next, we assume that the fields $X$ and $Y$ have the following form:
\begin{equation*}
X = X^{(K)} + \epsilon  \tilde X, \quad Y_{\mu} = Y^{(K)}_{\mu} + \epsilon  \tilde Y_{\mu}, 
\end{equation*}
where $\tilde X$ and $\tilde Y_{\mu}$ will be perturbations of the background solutions $X^{(K)}$ and $Y^{(K)}_{\mu}$ by a small factor $\epsilon$. 

Finally, after a tedious but straightforward procedure, we can linearize Eqs. (\ref{eq:casi_linear1}) and (\ref{eq:casi_linear2}) around the background solution $u^{(K)}=[X^{(K)}, Y^{(K)}_{\mu}]$ to obtain the following coupled system of equations:

\begin{equation}\label{ec:linearized_system}
\begin{split}
\partial_r  \tilde X   &=  \left( -\dfrac{3 }{2 r}   + \dfrac{\kappa_0}{r (X^{(K)})^2} \right) \tilde X + \dfrac{\alpha }{ \sqrt{2} \ r^2 } (   \overline{ \eth }  \tilde Y +  \eth \overline{ \tilde  Y }),   \\
\partial_r  \tilde Y_{1}   &=  -\dfrac{2 }{r}  \tilde Y_{1} + \dfrac{\alpha}{2 \sqrt{2}}\left( \dfrac{1}{2} + \dfrac{\kappa_{0}}{ (X^{(K)})^2 }  \right) \ \eth \tilde X,\\
\partial_r  \tilde Y_{2}   &=  \overline{ \partial_r  \tilde Y_{1} },
\end{split}
\end{equation}
where the last equatiation is just the complex conjugated of the first one. Clearly, this set of equations comprises a coupled linear system of PDEs, which, as we will demonstrate in the next section, could be rewritten in the general form of \Eqref{ec:general_linear_eq}. Consequently, it will be possible to use the algorithm presented in \Sectionref{sec:linear_case_algorithm} for its numerical solution.

To conclude this part, we define perturbations of the initial data $(h^{K}_{ab},\chi^{K}_{ab})$ as the pair 
$(h_{ab},\chi_{ab})$  such that 
\begin{equation}\label{definitionperturbations}
h_{ab} := h^{(K)}_{ab}, \quad \chi_{ab} := \chi^{(K)}_{ab} + \epsilon \tilde \chi_{ab},
\end{equation}
where $\tilde \chi_{ab}$ 
\begin{equation}\label{perturbationchi}
\tilde \chi_{ab} = \tilde Z n_{a}\otimes n_{b} + n_{a} \otimes \tilde Y_{b} +  n_{b} \otimes \tilde Y_{a} + \tilde K_{ab}
\end{equation}

with
\begin{eqnarray}\label{pertubed_kab}
\tilde K_{ab} &=& \dfrac{1}{2} \gamma_{ab} \tilde X, \quad \tilde Y_{a} = \tilde Y \mathbf{w}^2_a + \bar{ \tilde{Y} } \mathbf{w}^3_a,
\end{eqnarray}

and
\begin{equation}\label{pertubed_Z}
\tilde Z =  \dfrac{ \tilde Y  \bar Y^{(K)} +  \bar{\tilde Y}  \tilde Y^{(K)} }{ 2 X^{(K)} r^2 } - \dfrac{2 Y \bar Y}{X^{(K)} {}^2  r^2} \tilde X   -  \tilde X  X^{(K)},
\end{equation}

which corresponds to the linearization of \Eqref{eq:Constraint3}.

\subsection{Construction of the matricial system of spectral coefficients}\label{sec:constructionmatricialsystem}
For the sake of simplicity, from now on we will assume that the unknown functions $\tilde X$, $\tilde Y_{1}$ and $\tilde Y_{2}$ are axially symmetric, meaning they will not depend on the variable $\varphi$. Furthermore, since by the eth formalism $\tilde Y_{1}=\overline{ \tilde Y_{2}}$, we will just consider the equation for the first quantity and we will denoted by $\tilde Y=\tilde Y_{1}$.

Under the above assumptions, we can express these functions in terms of the \textbf{swsh} as:
\begin{equation}\label{eq:functions_decomposed_inS2}
\begin{split}
\tilde X &= \sum\limits_{l}   \hspace{0.1cm}_{0} x_{l}(r) \hspace{0.2cm}_{0}Y_{l} (\theta ) := \sum\limits_{l,0}   \hspace{0.1cm}_{0} x_{l0}(r) \hspace{0.2cm}_{0}Y_{l0} (\theta ,\varphi), \\
\tilde Y &=  \sum\limits_{l} \hspace{0.1cm}_{1} y_{l}(r) \hspace{0.2cm}_{1}Y_{l} (\theta) := \sum\limits_{l,0} \hspace{0.1cm}_{1} y_{l0}(r) \hspace{0.2cm}_{1}Y_{l0} (\theta,\varphi),
\end{split}
\end{equation}
where we have used the notation introduced in \cite{beyer2016numerical}:
\begin{equation*}
{}_{0} x_{l}(r)={}_{0} x_{l0}(r), \quad {}_{1} y_{l}(r)={}_{1} y_{l0}(r), \quad  {}_{0}Y_{l} (\theta) : = {}_{0}Y_{l0} (\theta, \varphi).
\end{equation*}
Note that the index $m$ is always zero due to the axial symmetry. As a result, it follows from \Eqsref{eq:eths} and  (\ref{ec:complex_conjugate_swsh}) that   we can write the system \Eqref{ec:linearized_system}  in the simple form  
\begin{equation} \label{ec:reducedperturbation} 
\begin{split}
\partial_r  \tilde X &= \tilde h^{(1)}  \tilde X + \tilde g^{(1)} \  \overline{ \eth }  \tilde Y , \\
\partial_r  \tilde Y &= \tilde h^{(2)} \tilde Y + \tilde f^{(2)}  \ \eth \tilde X,  
\end{split}
\end{equation}
where the functions $\tilde h^{(1)}$, $\tilde h^{(2)}$, $\tilde q^{(1)}$ and $\tilde q^{(2)}$ can be written in terms of the \textbf{swsh} as 
\begin{eqnarray}
\tilde h^{(1)}  &=& \sum\limits_{l}   \hspace{0.1cm}_{0} h^{(1)}_{l}(r) \hspace{0.2cm}_{0}Y_{l} (\theta,\varphi)  = - \dfrac{ \sqrt{4 \pi}(3 M + r)}{2 M r + r^2}  \ {}_0 Y_{0}(\theta ), \label{ec:spectral1} \\
\tilde h^{(2)}   &=& \sum\limits_{l}   \hspace{0.1cm}_{0} h^{(2)}_{l}(r) \hspace{0.2cm}_{0}Y_{l} (\theta )  =  -\dfrac{2 \sqrt{4 \pi} }{r} \  \ {}_0 Y_{0}(\theta ) ,\label{ec:spectral2} \\
\tilde g^{(1)} &=& \sum\limits_{l}   \hspace{0.1cm}_{0} g^{(1)}_{l}(r) \hspace{0.2cm}_{0}Y_{l} (\theta )  = 2 \dfrac{\sqrt{4 \pi}}{r^2} \sqrt{ \dfrac{1}{2} + \dfrac{M}{r} }   \ {}_0 Y_{0}(\theta ), \label{ec:spectral3} \\
\tilde f^{(2)}  &=& \sum\limits_{l}   \hspace{0.1cm}_{0} f^{(2)}_{l}(r) \hspace{0.2cm}_{0}Y_{l} (\theta )  =  \dfrac{\sqrt{4 \pi}(M + r)}{2 (2 M + r)} \sqrt{ \dfrac{1}{2} + \dfrac{M}{r} }  \  {}_0 Y_{0}(\theta ). \label{ec:spectral4}  
\end{eqnarray}
At this point we want to remark that only the first spectral coefficient  of the above functions basis is not zero. In order to write explicitly the above system as in the general form of \Eqref{ec:general_linear_eq}, we define the vector $\tilde u = [ \tilde X , \tilde Y ]$ and the following $2\times2$ diagonal and anti-diagonal matrices 
\begin{equation}\label{eq:definitions_h_f_g}
\mathcal{H} := \text{diag} \big[ \tilde h^{(1)} ,\tilde h^{(2)} \big], \ \mathcal{F} := \text{anti-diag} \big[ 0 , \tilde f^{(2)} \big], \ \mathcal{G} := \text{anti-diag}\big[ \tilde g^{(1)} ,0 \big].
\end{equation} 
Thus, we can compress the system \Eqref{ec:reducedperturbation} into the following equation matricial equation 
 \begin{equation}\label{ec:general_linear_eq_vectorial}
\partial_{r} \tilde u  =  \mathcal{F} \  \eth \tilde u  + \mathcal{G } \ \bar{\eth} \tilde u  + \mathcal{H}     \tilde u ,
\end{equation}
where $\partial_r \tilde u := ( \partial_r \tilde X , \partial_r \tilde Y )$,  $\eth \tilde u := (  \eth \tilde X , \eth \tilde Y )$ and $\bar \eth \tilde u := ( \bar \eth \tilde X , \bar \eth \tilde Y )$. Furthermore, by imposing the  
 the boundary conditions
\begin{equation}\label{ec:boundary_conditions_finalMatricial}
\begin{split} 
\tilde u(  r_0  )  &= \tilde u_0,\\  
\tilde u( r )  &= \mathcal{O}({r}^{-1}),
\end{split} 
\end{equation}
for some positive $r_0$, the matricial system \Eqref{ec:general_linear_eq_vectorial} takes a  form similar to that of \Eqref{ec:general_linear_eq} (with $q=0$). In what follows, we will show that the algorithm presented \Sectionref{sec:linear_case_algorithm} can be easily extended for treating this matricial system.

Considering the spectral decomposition of $\tilde X$ and $\tilde Y$ given in  \Eqref{eq:functions_decomposed_inS2}, we write $\tilde u$ as
\begin{equation}\label{ce:definitions_matricial1}
\begin{split}
\tilde u &= \sum\limits_{l_1}  \  \Upsilon_{l_1 } (\theta ) \ \mathcal{U}_{l_1 }(r),  \\
 \eth \tilde u &= \sum\limits_{l_1}  \  {}_{(+1)} \Upsilon_{l_1  } (\theta )\ 
  \ \mathcal{U}_{l_1 }(r),\\
 \bar \eth \tilde u &= \sum\limits_{l_1 }  \  {}_{(-1)}\Upsilon_{l_1  } (\theta )\ 
 \ \mathcal{U}_{l_1 }(r),
\end{split}
\end{equation}
where using \Eqref{eq:eths}, we have defined 
\begin{equation}\label{ce:definitions_matricial2}
\begin{split}
\mathcal{U}_{l_1 }(r) &:= (\ {}_{0} x_{l_1 }(r) \ , \ {}_{1}y_{l_1}(r) \ ), \\ \Upsilon_{l} (\theta ) &:= \text{diag} \big[ {}_{0}Y_{l} (\theta) \ , \ {}_{ 1}Y_{l} (\theta) \big] \\
{}_{(+1)}\Upsilon_{l } (\theta,\varphi) &:= \text{diag} \big[ S(l_1, 0, +1 ) \ {}_{1}Y_{l} (\theta) \ \ , \ S(l_1, 1,  +1 ) \ {}_{2}Y_{l} (\theta) \big], \\
 {}_{(-1)}\Upsilon_{l } (\theta,\varphi) &:= \text{diag} \big[   S(l_1,0,  -1 ) {}_{-1}Y_{l} (\theta) \ , \  S(l_1, 1, -1 ) \ {}_{0}Y_{l} (\theta ) \big]. 
\end{split}
\end{equation}
In what follows, we will refer to the term $\mathcal{U}_{l_1 }(r)$ as the vector-spectral coefficients.

Let $A$ be a a square matrix we denoted by $A=[a_{ij}]$. Further, we will write the product of matrices  as $A B = [ a_{ij} ]  [ b_{jk} ] =  \big[ \sum_j  a_{ij} b_{jk}   \big]$ and the  integration over components of matrices   as
\begin{equation*}
\int \limits^{b}_{a} A \ dx = \int \limits^{b}_{a} \big[ a_{ij} \big] \ dx = \Big[ \int \limits^{b}_{a} a_{ij} \ dx \Big] .
\end{equation*}
Then, using the above notation, we extend the inner product $ \langle \ , \  \rangle$ used in \Sectionref{sec:linear_case_algorithm} to the components of matrices as follows  
\begin{equation}\label{inner:product:ecuation}
\begin{split}
\langle \  A  ,  B  \  \rangle & = \langle \  [ a_{ij} ]  ,  [ b_{jk} ]  \  \rangle  :=  
\int \limits_{\mathbb{S}^2}
 \overline{ [ a_{ij} ] } [ b_{jk} ] \ d \Omega  
 = \int \limits_{\mathbb{S}^2}
\Big[ \sum_j \bar a_{ij}  b_{jk} \Big] \ d \Omega \\
&= 
\Big[  \int \limits_{\mathbb{S}^2} \sum_j \bar a_{ij}  b_{jk} \ d \Omega \Big] 
 = \Big[ \sum_j \int \limits_{\mathbb{S}^2}   \bar a_{ij}   b_{jk} d \Omega   \Big]  = \Big[ \sum_j \langle  a_{ij} , b_{jk}   \rangle  \Big] .
\end{split}  
\end{equation}
Note that because of the linearity of the integration, it is clear that $\langle \  A  ,  B +C  \  \rangle = \langle \  A  ,  B  \  \rangle = \langle \  A  ,  C  \  \rangle$. Furthermore, it follows that the special product $\langle \  A  ,  B  \ v \  \rangle$, with $v$ being a vector whose components only depend on $r$, must be equal to $\langle \  A  ,  B  \ v \  \rangle = \langle \  A  ,  B  \  \rangle \ v$.
 
Using the above inner product for matrices, we can project the equation \Eqref{ec:general_linear_eq_vectorial} to the basis $\Upsilon(\theta)_l$ to obtain
\begin{equation}\label{ec:proyecting_final_equation}
\langle \ \Upsilon_{l} (\theta ) , \partial_r  \tilde u    \ \rangle =  \langle \ \Upsilon_{l} (\theta) , \mathcal{H} \tilde u    \ \rangle + \langle \ \Upsilon_{l} (\theta) , \ \mathcal{F} \ \eth \tilde u  \ \rangle + \langle \ \Upsilon_{l} (\theta) ,  \ \mathcal{G} \ \bar \eth \tilde u  \ \rangle.
\end{equation}
Clearly, from the left hand side of the equation and \Eqref{ce:definitions_matricial1} we find that
\begin{eqnarray}\label{ec:system0}
\langle \ \Upsilon_{l} (\theta ) , \partial_r  \tilde u    \ \rangle &  =&
\langle  \ \Upsilon_{l} (\theta) ,  \sum\limits_{l}  \  \Upsilon_{l_1 } (\theta) \  \partial_r \ \mathcal{U}_{l}(r)  \  \rangle , \nonumber \\
 &=& \sum\limits_{l}  \langle  \ \Upsilon_{l} (\theta) ,   \Upsilon_{l_1} (\theta,\varphi) \      \rangle \  \partial_r \ \mathcal{U}_{l}(r)  ,\nonumber \\
 &=& \sum\limits_{l}    \text{diag} \Big[   \langle  {}_{0}Y_{l} (\theta )  ,   {}_{0}Y_{l_1} (\theta)   \rangle   ,   \langle   {}_{1}Y_{l} (\theta)  ,   {}_{1}Y_{l_1} (\theta)   \  \rangle    \Big] \  \partial_r \ \mathcal{U}_{l    } (r)  ,\nonumber \\
& = &  \sum\limits_{l}    \ \text{diag} \Big[ \ \delta_{l_1 l}  , \  \delta_{l_1 l}   \Big]   \partial_r \ \mathcal{U}_{l_1} (r)  =    \partial_r \ \mathcal{U}_{l }(r) .
\end{eqnarray}
On the other hand, from \Eqsref{eq:definitions_h_f_g} and the formula for the  $\mathcal{C}$-functions \Eqref{ec:C-equation},  the first term of the right hand side 
becomes
\begin{eqnarray}\label{ec:system1}
\langle \ \Upsilon_{l} (\theta) , \mathcal{H} \tilde u    \ \rangle &  =&
\langle  \ \Upsilon_{l} (\theta) , \mathcal{H}  \sum\limits_{l}  \  \Upsilon_{l_1} (\theta) \ \mathcal{U}_{l}(r)  \  \rangle  
 = \sum\limits_{l_1}  \langle  \ \Upsilon_{l_1 } (\theta) , \mathcal{H} \ \Upsilon_{l_1} (\theta) \   \rangle \  \mathcal{U}_{l }(r)  \nonumber ,\\
 &=& \sum\limits_{l_1}  \text{diag} \Big[   \langle  {}_{0}Y_{l} (\theta) ,  \tilde h^{(1)}  \ {}_{0}Y_{l_1 } (\theta)   \rangle   ,     {}_{1}Y_{l} (\theta) , \tilde h^{(2)} \ {}_{1}Y_{l} (\theta)   \  \rangle    \Big] \  \mathcal{U}_{l_1} (r)  ,\nonumber \\
& = & \sum\limits_{l_1}    \ \text{diag} \Big[ \ \mathcal{C}_{(0) l_1 0,0  l 0} ( \{  {}_{0}h^{(1)}_{l_2 0}(r) \}  )  \ , \  \mathcal{C}_{(1) l_1 0,1 l 0} (  \{  {}_{0 }h^{(2)}_{l_2 0}(r) \}  )  \     \Big] \ \mathcal{U}_{l_1} (r) ,\nonumber \\
& = &  \text{diag} \Bigg[ {}_{0 }h^{(1)}_{0}(r)  \ , \  {}_{0 }h^{(2)}_{0}(r)   \Bigg] \ \mathcal{U}_{l} (r) .
\end{eqnarray}
Note that in this case, the $\mathcal{C}$-functions do not generate a coupling between the vector-spectral coefficients $\mathcal{U}l$ because all the spectral coefficients ${}{0}h^{(1)}l$ and ${}{0}h^{(2)}_l$ vanish for l>0l>0l>0 (refer to \Eqsref{ec:spectral1} and (\ref{ec:spectral2})). Following a similar procedure, we derive the following expressions from the second and third terms of \Eqref{ec:proyecting_final_equation}:
\begin{eqnarray}\label{ec:system2}
\langle \ \Upsilon_{l} (\theta) , \ \mathcal{F} \ \eth \tilde u  \ \rangle &  =&
\langle  \ \Upsilon_{l} (\theta) , \ \mathcal{F}  \sum\limits_{l_1}  \  {}_{(+1)}\Upsilon_{l_1} (\theta) \ \mathcal{U}_{l_1}(r)  \  \rangle,  \nonumber \\
 &=& \sum\limits_{l_1 }  \langle  \ \Upsilon_{l_1 } (\theta) , \mathcal{F} \ {}_{(+1)}\Upsilon_{l_1} (\theta) \   \rangle \  \mathcal{U}_{l_1 }(r) , \nonumber \\
 &=& \sum\limits_{l_1}    \text{anti-diag} \Big[ \ 0 \ , \langle   {}_{1}Y_{l} (\theta) ,  \tilde f^{(2)} \  S(l_1,0,+1) \ {}_{1}Y_{l_1} (\theta)    \rangle \   \Big] \  \mathcal{U}_{l_1 } (r)  ,\nonumber \\
& = & \sum\limits_{l_1}    \ \text{anti-diag} \Big[ \ 0  \ , \mathcal{C}_{(1) l_1 0,1 l 0} ( \{  {}_{0}f^{(2)}_{l_2 0}(r) \}  )  \  \mathcal{S}(l_1,0, +1) \  \Big] \ \mathcal{U}_{l_1 } (r) ,\nonumber \\
& = &  \text{anti-diag} \Bigg[ \ 0 , {}_{0}f^{(2)}_{0 }(r) \ \mathcal{S}(l,0, +1) \    \ \Bigg] \ \mathcal{U}_{l} (r)  ,
\end{eqnarray}

\begin{eqnarray}\label{ec:system3}
\langle \ \Upsilon_{l} (\theta) ,  \ \mathcal{G} \ \bar \eth \tilde u  \ \rangle &  =&
\langle  \ \Upsilon_{l} (\theta) , \ \mathcal{G} \sum\limits_{l_1}  \  {}_{(-1)}\Upsilon_{l_1} (\theta) \ \mathcal{U}_{l_1}(r)  \  \rangle = \sum\limits_{l_1 }  \langle  \ \Upsilon_{l_1 } (\theta) , \mathcal{G} \ {}_{(-1)}\Upsilon_{l_1} (\theta) \   \rangle \  \mathcal{U}_{l_1 }(r) , \nonumber \\
 &=& \sum\limits_{l_1}   \text{anti-diag} \Big[ \ \langle   {}_{1}Y_{l} (\theta) ,  \tilde g^{(1)} \  S(l_1,1,-1) \ {}_{1}Y_{l_1} (\theta)    \rangle \ , \ 0 \   \Big] \  \mathcal{U}_{l_1 } (r) , \nonumber \\
& = & \sum\limits_{l_1}    \ \text{anti-diag} \Big[ \mathcal{C}_{(1) l_1 0,1 l 0} ( \{  {}_{0}g^{(1)}_{l_2 0}(r) \}  )  \ S(l_1,1,-1) , 0 \Big] \ \mathcal{U}_{l_1 } (r), \nonumber  \\
& = &  \text{anti-diag} \Bigg[     {}_{0}g^{(1)}_{0}(r) \ S(l,1,-1) , \ 0 \   \Bigg] \ \mathcal{U}_{l } (r) .
\end{eqnarray}
Finally, substituting \Eqsref{ec:system0}-(\ref{ec:system3}) into \Eqref{ec:proyecting_final_equation},  we obtain the matricial system for the vector-spectral coefficients
\begin{equation}\label{ec:final_matricial_system}
\partial_r \mathcal{U}_{l }(r)  = A_l(r)  \ \mathcal{U}_{l }(r), \quad \text{ for } l=0,1,2...,
\end{equation}
where 
\begin{equation*}
A_l(r):=\begin{bmatrix}
{}_{0}h^{(1)}_{0}(r) & {}_{0}g^{(1)}_{0}(r) \  S(l,1,-1)  \\
{}_{0}f^{(2)}_{0}(r) \ S(l,0,+1) & {}_{0}h^{(2)}_{0}(r) 
\end{bmatrix},
\end{equation*}
and subject to the boundary conditions given by  \Eqref{ec:boundary_conditions_finalMatricial} for each vector-spectral coefficients, that is,
\begin{equation}\label{asymtotic_decay_condition}
\begin{split}
\mathcal{U}_l(r_0) &= \mathcal{U}^{(0)}_l , \\
\mathcal{U}_l(r)   &= \mathcal{O}({r}^{-1}), 
\end{split}  
\end{equation}
for $\mathcal{U}^{(0)}_l$ being the vector-spectral coefficients of the initial data $\tilde u_0$. Note that we have obtained a system that does not contain any coupling between the vector-spectral coefficients $\mathcal{U}_l$. As we pointed out just after \Eqref{ec:system1}, this is because of only the spectral coefficients  ${}_{0}h^{(1)}_0$, ${}_{0}h^{(2)}_0$, ${}_{0}f^{(2)}_0$ and ${}_{0} g^{(1)}_0$ are different from zero.

\subsection{Numerical solutions}

To commence, let us establish a mesh $(r_i,\theta_j,\phi_k)$ on $[r_0,\infty) \times \mathbb{S}^2$ comprising $N_r \times N_\theta \times N_\varphi$ points, as detailed in \Sectionref{sec:subsectionspectraldecomposition}. It is important to note that as the functions $\tilde h^{(1)}, \tilde h^{(2)}, \tilde g^{(1)}$, and $\tilde f^{(2)}$ can be readily decomposed into the \textbf{swsh} basis (refer to \Eqsref{ec:spectral1}--\ref{ec:spectral4}), we solely require angular discretization to ultimately derive the numerical solutions of $\tilde X$ and $\tilde Y$ from their vector-spectral coefficients.

To rewrite the system \Eqref{ec:final_matricial_system} in its variational form, we assume the solution exists and is unique within a finite-dimensional vector space $\mathscr{V} \subseteq L^2([a,\infty)) \times L^2([a,\infty))$. Consequently, it must hold for any test-column vector $V$ in $\mathscr{V}$ that:

\begin{equation}\label{ec:variationalform21}
\int_{r_0}^{\infty}  V   \cdot  \Big(  \partial_r \mathcal{U}_{l }(r)  - A_l(r) \ \mathcal{U}_{l}(r) \Big) \ dr = 0 ,
\end{equation}
where $\cdot$ denotes the standard inner product of two-dimensional Euclidean vectors. To derive a system of algebraic equations for the unknown functions ${}_{0}x_{l}(r)$ and ${}_{0}y_{l}(r)$, we need to select linearly independent test-column vectors $V$ to generate two independent variational equations from \Eqref{ec:variationalform21}. This can be achieved easily, for example, by choosing $V_1:=(v_1,0)$ and $V_2:=(0,v_2)$. Furthermore, akin to \Sectionref{sec:constructionmatricialsystem}, by introducing the test diagonal matrix
\begin{equation*}
\mathcal{V} := \text{diag}\big[ \  v_1 ,v_2  \   \big],
\end{equation*}
we can compress the two independent equations into the following  matricial equation 
\begin{equation}\label{ec:variationalform2}
\int_{r_0}^{\infty}  \mathcal{V} \   \Big(  \partial_r \mathcal{U}_{l }(r)  - A_l(r) \ \mathcal{U}_{l}(r) \Big) \ dr = 0,
\end{equation}
Next, similarly as in \Sectionref{sec:inifiniteelementmethod}, we use the Zienkiewicz coordinate transformation  (\Eqref{ec:coordinate_transform_fem}) over the radial points $r_i$ with pole $r_p=0$,  in order to find  $N_r+1$-points  $\xi_0, ..., \xi_{N_r}$,  in the interval $[-1,1]$ with $\xi_0=-1$ and $\xi_{N_r}=1$. Additionally, we can write the unknown functions ${}_{0}x_{l}(r) $ and $ {}_{0}y_{l}(r)$ in terms of  the  compactly supported polynomials  $\psi_i(\xi)$ defined on $[-1,1]$ (see \Eqref{ec:cspolynomials}) by writing $\mathcal{U}_l(r)$ as follows
\begin{equation}
\mathcal{U}_l (\xi) = \sum_{i=1}^{N_r} \Psi_i(\xi) \cdot \mathcal{U}_l (\xi_i), 
\end{equation}
where  
\begin{equation*}
 \Psi_i(\xi) := \text{diag}\big[ \    \psi_i(\xi)   ,   \psi_i(\xi)   \   \big].
\end{equation*}
Substituting the above into the variational form \Eqref{ec:variationalform2} we obtain
\begin{equation*} 
 \sum_{i=1}^{N_r} \int_{-1}^{1}  \mathcal{V}   \  \Big(   \partial_r \Psi_i(\xi)  - A_l(\xi) \cdot \Psi_i(\xi)  \Big) \  \mathcal{U}_l (\xi_i)  \ J(r;\xi) \ d\xi = 0.
\end{equation*}
Finally, choosing $N_{r}$ test diagonal matrices as $\mathcal{V}_j :=   \Psi_j(\xi)$, we obtain  the following linear system of equations
\begin{equation}\label{ec:blocksystem}
\mathcal{A}^{(l)}  \ \mathscr{U}^{(l)}  = 0,
\end{equation}
with $\mathscr{U}_l$ being a vector of $N_r$ blocks defined by
\begin{equation*}
\mathscr{U}^{(l)}_{i}  := \mathcal{U}_l(\xi_i), \quad i=1,...,N_r,\\
\end{equation*}
and $\mathcal{A}$  a  matrix of  $N_r \times N_r$ blocks given by the  $2\times2$ matrices
\begin{equation*} 
\mathcal{A}^{(l)}_{ji} =  \int_{-1}^{1}  \partial_r \Psi_j(\xi)  \   \Big(   \partial_r \Psi_i(\xi)  - A_l(\xi) \cdot \Psi_i(\xi)  \Big)  \ J(r;\xi) \ d\xi .  
\end{equation*}
Because of $\mathcal{A}_l$ is a square matrix of $2N_r \times 2N_r$ entries and $\mathscr{U}^{(l)}$ is a vector of $2N_r$ components,  the system \Eqref{ec:blocksystem} is  a coupled linear system of $2N_r \times 2N_r$ equations, which we solve numerically as follows. 

First, we choose $N_r= 200$ points in $[-1,1]$ equally spaced. Second, we impose the following   boundary conditions over the spectral coeficients $\mathcal{U}_1$ and $\mathcal{U}_2$:
\begin{equation*}
\begin{split}
\mathcal{U} _1(-1) &= \big[ 0.05 , -0.04 \big],  \\
\mathcal{U} _1(\ 1) &=\big[ 0 , 0  \big], 
\end{split}  
\end{equation*}
and 
\begin{equation*}
\begin{split}
\mathcal{U}_2(-1) &=  \big[ 0.08 , 0.07 \big],  \\
\mathcal{U}_2(\ 1) &=\big[ 0 , 0  \big]. 
\end{split}  
\end{equation*}
For the rest of the coefficients $\mathcal{U}_l$, specifically for $2 < l \leq L$, where $L$ represents the band limit (as discussed in \Sectionref{sec:subsectionspectraldecomposition}), we will impose the condition that these coefficients vanish at both boundary points. Additionally, as a third step, we will utilize $\psi(\xi)$ as second-order interpolating Lagrangian polynomials, as elaborated in references such as \cite{gockenbach2006understanding}.
\begin{figure}[t]
    \centering
     \includegraphics[scale=0.8]{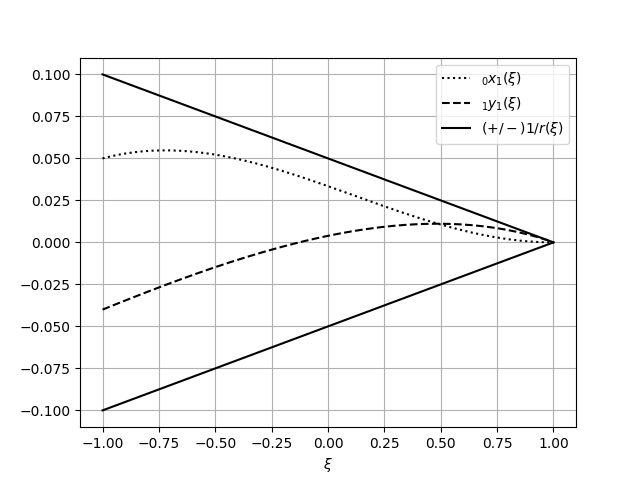}  
     \caption{Numerical solution of ${}_{0} x(\xi)_{1}$ and ${}_{1} y(\xi)_{1}$ in terms of the variable $\xi$.} \label{fig:figure3}    
    \centering
      \includegraphics[scale=0.8]{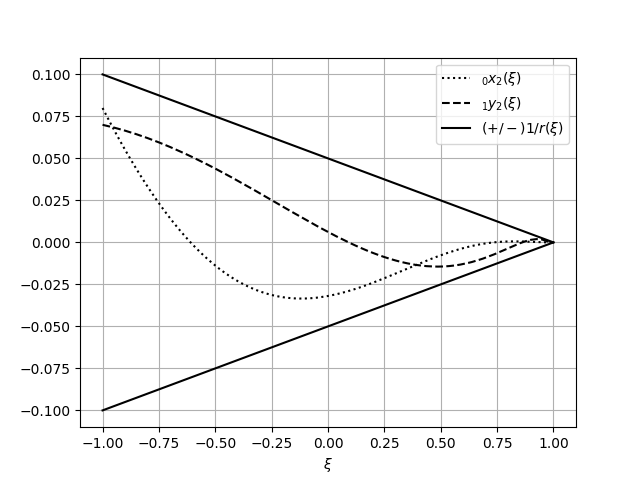}   
     \caption{Numerical solution of ${}_{0} x(\xi)_{2}$ and ${}_{1} y(\xi)_{2}$ in terms of the variable $\xi$.} \label{fig:figure4}
\end{figure} 

Since the linear equations \Eqref{ec:blocksystem} are decoupled for each $l$, we can solve them independently for each spectral coefficient $\mathcal{U}_l$ using various linear algebra software. In our specific case, we implemented a routine in the \textbf{Python} language (refer to \cite{zelle2004python}) to solve these linear systems of equations, utilizing the \texttt{linalg} module within the \texttt{scipy} package. This package can be freely downloaded from its official website at \url{https://www.scipy.org/}.


Because of the boundary conditions for $\mathcal{U}_{l}$ with $l = 3, ..., L$ , we obtained that they vanish in the all domian. On the other hand, in \Figref{fig:figure3} and \Figref{fig:figure4}, we show the numerical solutions for $\mathcal{U}_1$ and $\mathcal{U}_2$, respectively. Note that in both figures, we have included the lines $1/r(\xi)$  and $-1/r(\xi)$ (both in bold-line style) for remarking that the components of the numerical solution $\mathcal{U}_l$, with $l=1,2$,  satisfy the asymptotic decay  condition of \Eqsref{asymtotic_decay_condition}. 

\subsection{Asymptotic behavior}

We devote this last subsection to establish whether the numerical solutions that we found above are asymptotically flat or not. Do to so, we first note that by a straightforward computation, it can be easily obtained that any tensor of the form 
\begin{eqnarray}\label{ec:formula_general_equation}
T_{ab}&=&   \mathcal{O}({r}^{-2}) \mathbf{w}^0_a \otimes \mathbf{w}^0_b +   \mathcal{O}({r}^{-1})   \mathbf{w}^0_a \otimes  ( \mathbf{w}^1_b -  \mathbf{w}^2_b ) +  \mathcal{O}({r}^{0})   \mathbf{w}^1_a \otimes  \mathbf{w}^2_b 
\end{eqnarray}
in the  standard spherical coordinates, takes the following form in standard Cartesian coordinates $(\tilde x^1 , \tilde x^2, \tilde x^3 )$:
\begin{eqnarray}\label{ec:formula_general_equation_cartesian}
T_{ab}&=&   \sum_{i,j=1}^{3} \mathcal{O}({r}^{-2}) \df \tilde x^{i}_a \otimes \df \tilde x^{j}_a ,
\end{eqnarray}
where  the $\df \tilde x^{i}_a$ are the dual basis covectors associated to the coordinate frame, i.e., $\df \tilde x^{i}_a = \nabla_a \tilde x^{i}$. 

Second, we recall that in general, it can be proved that the initial data $(h_{ab},\chi_{ab})$ given in \Eqref{Kerr_metric_cauchy} that describe the $t$-constant Cauchy surfaces of the Kerr spacetime, satisfy the definition \Eqref{def:definition_asymptotically_flat} in Cartesian coordinates (see \cite{cook1997well}). 

With the above in mind,  and considering that we are computing  perturbations $(h_{ab},\chi_{ab})$ of the form of \Eqref{definitionperturbations}, we only have to determine whether the tensor $\tilde \chi_{ab}$ defined by \Eqref{perturbationchi} has or not the form  of \Eqref{ec:formula_general_equation}. Since this tensor is just the composition of three terms, namely; $n_{a} \otimes \tilde Y_{b}$, $\tilde K_{ab}$ and $\tilde Z n_{a} \otimes n_{b}$, we will examine their behavior separately. 

To begin with, we note that because of $\alpha= \mathcal{O}({r}^{0})$ (see \Eqsref{alpfaybetaforkerr}), it follows from \Eqref{ec:normalVector_frame} that $n_{a} =  \mathcal{O}({r}^{0})  \mathbf{w}^0_a$. Additionally, since we found numerical solutions  such that $\mathcal{U}_l = \mathcal{O}({r}^{-1})$ for $l=1,2$, it follows that $\tilde Y=\mathcal{O}({r}^{-1})$. Thus, combining these two facts and the second equation of \Eqsref{pertubed_kab} we obtain that 
\begin{equation}\label{eq:finalequation1}
n_{a} \otimes \tilde Y_{b} =  \mathcal{O}({r}^{-1})  \mathbf{w}^0_a \otimes \mathbf{w}^2_b+ \mathcal{O}({r}^{-1}) \mathbf{w}^0_a \otimes \mathbf{w}^3_b.
\end{equation}

\begin{figure}[t]
    \centering
     \includegraphics[scale=0.8]{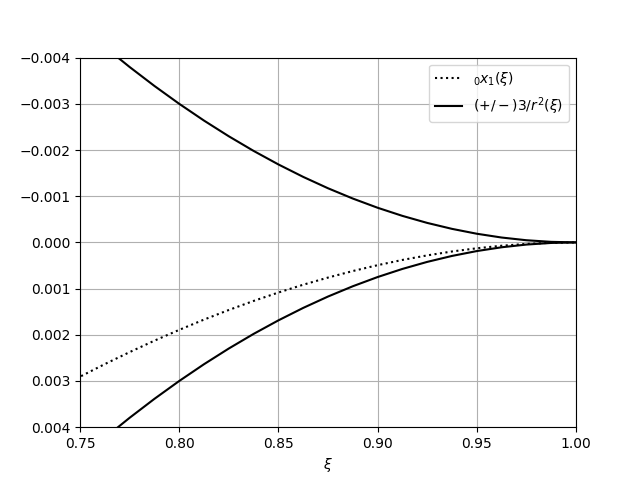}  
     \caption{Numerical solution of ${}_{0} x(r)_{1}$  between the plot of $3/r(\xi)$ and $-3/r(\xi)$.} \label{fig:figure7}    
    \centering
      \includegraphics[scale=0.8]{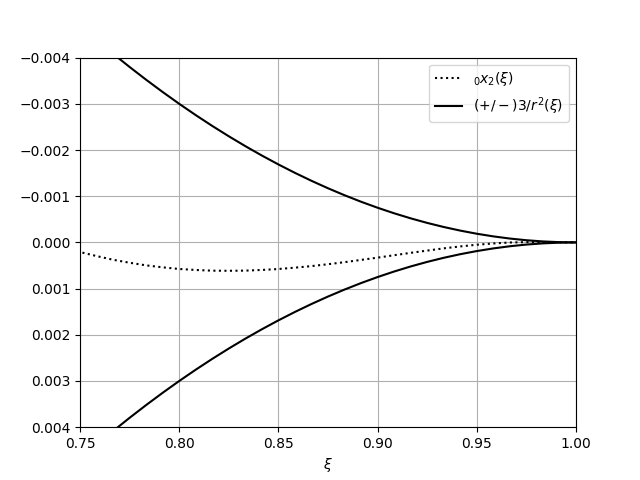}   
     \caption{Numerical solution of ${}_{0} x(r)_{2}$ between the plot of $3/r(\xi)$ and $-3/r(\xi)$.} \label{fig:figure8}
\end{figure} 

Before we proceed to examine the behavior of $\tilde K_{ab}$, we first  consider \Figref{fig:figure7} and \Figref{fig:figure8}. There it is displayed the behavior of ${}_{0}x_1(\xi)$ and ${}_{0}x_2(\xi)$  in the last quarter of the numerical domain; that is $[0.75, 1]$. In each of these figures, we also include the plots of the functions $3/r^2(\xi)$ and $-3/r^2(\xi)$, which  suggest that both ${}_{0}x_1(\xi)$ and ${}_{0}x_2(\xi)$ are $\mathcal{O}({r}^{-2})$. Therefore,  this fact in combination with \Eqref{pertubed_kab} leads to
\begin{equation}\label{eq:finalequation2}
\tilde K_{ab} = \dfrac{r^2 }{2} \mathring{\gamma}_{ab} \tilde X  = \mathcal{O}({r}^{0}) \mathbf{w}^1_a \otimes \mathbf{w}^2_b. 
\end{equation}
Furthermore, from \Eqref{pertubed_Z} we obtain after some computations that $\tilde Z = \mathcal{O}({r}^{-2})$, which implies
\begin{equation}\label{eq:finalequation3}
\tilde Z n_{a} \otimes n_{b} =  \mathcal{O}({r}^{-2}).
\end{equation}

Finally, by substituting \Eqsref{eq:finalequation1}-(\ref{eq:finalequation3}) in \Eqref{perturbationchi} we obtain the desired result,  that is; the tensor $\tilde\chi_{ab}$ can be written as \Eqref{ec:formula_general_equation}, thus, it can take the form of \Eqref{ec:formula_general_equation_cartesian} in standard Cartesian coordinates. 

As a result of the above, we conclude that the perturbed initial data $(h_{ab},\chi_{ab})$ that we just found numerically in the last subsection is  asymptotically flat.

\section{Conclusions}\label{section:quinta_seccion} 

Motivated by R\'acz's hyperbolic formulation of the constraint equations, this work introduces a spectral-infinite element method for numerically solving partial differential equations in unbounded domains of the form $[r_0,\infty)\times \mathbb{S}^2$. This numerical approach is based on the pseudo-spectral method introduced by Beyer et al. in \cite{beyer2017asymptotics} (also see \cite{beyer2014numerical,beyer2016numerical} for other applications of this method), which deals with evolution-like problems having the spatial topology of $\mathbb{S}^2$, utilizing the eth-operators and the \textbf{swsh}.

While the aforementioned method employs an explicit Runge-Kutta approach to integrate the equations up to a large value along the radial coordinate, the approach presented in this work utilizes an infinite-element method, enabling the numerical integration of equations in the unbounded domain $[r_0,\infty)\times \mathbb{S}^2$. The main advantage of this new approach over the method in \cite{beyer2017asymptotics} lies in its capability to address unbounded domains. However, due to the complexity of the infinite-element aspect, it might pose challenges in implementing for solving non-linear PDEs.

As an application, the spectral-infinite element method is employed in R\'acz's hyperbolic formulation to compute asymptotically flat perturbations of a Kerr black hole with small angular momentum. It is important to note that for the sake of method robustness and brevity, we utilized this approach to solve a simplified system of linear PDEs (perturbation equations). Nonetheless, it is highly conceivable and promising to extend this numerical infrastructure to tackle more intricate systems of PDEs, which we leave for future research endeavors.


\section*{Acknowledgments}
This work was supported by Patrimonio Autónomo - Fondo Nacional de Financiamiento para la Ciencia, la Tecnología y la Innovación Francisco José de Caldas (MINCIENCIAS - COLOMBIA) Grant No. 110685269447 RC-80740-465-202, projects 69723 and 69553.

\addcontentsline{toc}{section}{References}
\bibliography{bibliography}

\begin{thebibliography}{10}

\bibitem{alcubierre2008introduction}
M.~Alcubierre.
\newblock {\em Introduction to 3+1 numerical relativity}, volume 140.
\newblock Oxford University Press, 2008.

\bibitem{bartnik2004constraint}
R.~Bartnik and J.~Isenberg.
\newblock The constraint equations.
\newblock In {\em The Einstein equations and the large scale behavior of
  gravitational fields}, pages 1--38. Springer, 2004.

\bibitem{baumgarte2010numerical}
T.~W. Baumgarte and S.~L. Shapiro.
\newblock {\em Numerical relativity: solving Einstein's equations on the
  computer}.
\newblock Cambridge University Press, 2010.

\bibitem{beyer2014numerical}
F.~Beyer, B.~Daszuta, J.~Frauendiener, and B.~Whale.
\newblock Numerical evolutions of fields on the 2-sphere using a spectral
  method based on spin-weighted spherical harmonics.
\newblock {\em Classical and Quantum Gravity}, 31(7):075019, 2014.

\bibitem{beyer2016numerical}
F.~Beyer, L.~Escobar, and J.~Frauendiener.
\newblock {Numerical solutions of Einstein’s equations for cosmological
  spacetimes with spatial topology $\mathbb{S}^3$ and symmetry group U(1)}.
\newblock {\em Physical Review D}, 93(4):043009, 2016.

\bibitem{beyer2017asymptotics}
F.~Beyer, L.~Escobar, and J.~Frauendiener.
\newblock Asymptotics of solutions of a hyperbolic formulation of the
  constraint equations.
\newblock {\em Classical and Quantum Gravity}, 34(20):205014, 2017.

\bibitem{brenner2007mathematical}
S.~Brenner and R.~Scott.
\newblock {\em The mathematical theory of finite element methods}, volume~15.
\newblock Springer Science \& Business Media, 2007.

\bibitem{butcher2008numerical}
J.~C. Butcher and N.~Goodwin.
\newblock {\em Numerical methods for ordinary differential equations},
  volume~2.
\newblock Wiley Online Library, 2008.

\bibitem{cook2000initial}
G.~B. Cook.
\newblock Initial data for numerical relativity.
\newblock {\em Living Reviews in Relativity}, 3(1):5, 2000.

\bibitem{cook1997well}
G.~B. Cook and M.~A. Scheel.
\newblock Well-behaved harmonic time slices of a charged, rotating, boosted
  black hole.
\newblock {\em Physical Review D}, 56(8):4775, 1997.

\bibitem{dain2001asymptotically}
S.~Dain and H.~Friedrich.
\newblock Asymptotically flat initial data with prescribed regularity at
  infinity.
\newblock {\em Communications in Mathematical Physics}, 222(3):569--609, 2001.

\bibitem{ern2013theory}
A.~Ern and J.-L. Guermond.
\newblock {\em Theory and practice of finite elements}, volume 159.
\newblock Springer Science \& Business Media, 2013.

\bibitem{escobar2016studies}
L.~Escobar.
\newblock {\em {Studies of spacetimes with spatial topologies $\mathbb{S}^3$
  and $\mathbb{S}^3 \times \mathbb{S}^2$}}.
\newblock PhD thesis, University of Otago, 2016.

\bibitem{garat2000nonexistence}
A.~Garat and R.~H. Price.
\newblock Nonexistence of conformally flat slices of the {K}err spacetime.
\newblock {\em Physical Review D}, 61(12):124011, 2000.

\bibitem{gerdes2000review}
K.~Gerdes.
\newblock A review of infinite element methods for exterior {H}elmholtz
  problems.
\newblock {\em Journal of Computational Acoustics}, 8(01):43--62, 2000.

\bibitem{geroch1972structure}
R.~Geroch.
\newblock Structure of the gravitational field at spatial infinity.
\newblock {\em Journal of Mathematical Physics}, 13(7):956--968, 1972.

\bibitem{gilbarg2015elliptic}
D.~Gilbarg and N.~S. Trudinger.
\newblock {\em Elliptic partial differential equations of second order}.
\newblock Springer, 2015.

\bibitem{gockenbach2006understanding}
M.~S. Gockenbach.
\newblock {\em Understanding and implementing the finite element method},
  volume~97.
\newblock Siam, 2006.

\bibitem{huffenberger2010fast}
K.~M. Huffenberger and B.~D. Wandelt.
\newblock Fast and exact spin-s spherical harmonic transforms.
\newblock {\em The Astrophysical Journal Supplement Series}, 189(2):255, 2010.

\bibitem{isenberg2014initial}
J.~Isenberg.
\newblock The initial value problem in general relativity.
\newblock In {\em Springer handbook of spacetime}, pages 303--321. Springer,
  2014.

\bibitem{lehner2001numerical}
L.~Lehner.
\newblock Numerical relativity: a review.
\newblock {\em Classical and Quantum Gravity}, 18(17):R25, 2001.

\bibitem{matzner1998initial}
R.~A. Matzner, M.~F. Huq, and D.~Shoemaker.
\newblock Initial data and coordinates for multiple black hole systems.
\newblock {\em Physical Review D}, 59(2):024015, 1998.

\bibitem{nakahara2003geometry}
M.~Nakahara.
\newblock {\em Geometry, topology and physics}.
\newblock CRC Press, 2003.

\bibitem{newman1966note}
E.~T. Newman and R.~Penrose.
\newblock Note on the bondi-metzner-sachs group.
\newblock {\em Journal of Mathematical Physics}, 7(5):863--870, 1966.

\bibitem{penrose1984spinors}
R.~Penrose and W.~Rindler.
\newblock {\em Spinors and space-time: Volume 1, Two-spinor calculus and
  relativistic fields}, volume~1.
\newblock Cambridge University Press, 1984.

\bibitem{racz2014cauchy}
I.~R{\'a}cz.
\newblock Cauchy problem as a two-surface based ‘geometrodynamics’.
\newblock {\em Classical and Quantum Gravity}, 32(1):015006, 2014.

\bibitem{racz2014bianchi}
I.~R{\'a}cz.
\newblock Is the bianchi identity always hyperbolic?
\newblock {\em Classical and Quantum Gravity}, 31(15):155004, 2014.

\bibitem{racz2015constraints}
I.~R{\'a}cz.
\newblock Constraints as evolutionary systems.
\newblock {\em Classical and Quantum Gravity}, 33(1):015014, 2015.

\bibitem{shibata2015numerical}
M.~Shibata.
\newblock {\em Numerical Relativity}, volume~1.
\newblock World Scientific, 2015.

\bibitem{stoer2013introduction}
J.~Stoer and R.~Bulirsch.
\newblock {\em Introduction to numerical analysis}, volume~12.
\newblock Springer Science \& Business Media, 2013.

\bibitem{Wald:1984un}
R.~M. Wald.
\newblock {\em General Relativity}.
\newblock University of Chicago Press, 1984.

\bibitem{wriggers2008nonlinear}
P.~Wriggers.
\newblock {\em Nonlinear finite element methods}.
\newblock Springer Science \& Business Media, 2008.

\bibitem{zelle2004python}
J.~M. Zelle.
\newblock {\em Python programming: {A}n introduction to computer science}.
\newblock Franklin, Beedle \& Associates, Inc., 2004.

\bibitem{zienkiewicz1983novel}
O.~Zienkiewicz, C.~Emson, and P.~Bettess.
\newblock A novel boundary infinite element.
\newblock {\em International Journal for Numerical Methods in Engineering},
  19(3):393--404, 1983.

\bibitem{zienkiewicz2000finite}
O.~C. Zienkiewicz, R.~L. Taylor, and J.~Z. Zhu.
\newblock {\em The finite element method: {I}ts basis and fundamentals}.
\newblock Elsevier, 2005.

\end{thebibliography}

\end{document}